\documentclass[aps,prl,twocolumn,superscriptaddress,footinbib,longbibliography]{revtex4-1} 

\usepackage{longtable}
\usepackage{morefloats}
\usepackage[dvips]{graphicx}
\usepackage{color}
\usepackage{epsfig,graphicx,amsfonts,amsbsy}
\usepackage{amsmath,amsfonts,amsthm,amssymb}
\usepackage{appendix}
\usepackage{makeidx}
\usepackage{url}
\usepackage{verbatim}
\usepackage[bookmarksnumbered,pdfpagelabels=true,plainpages=false,colorlinks=true,linkcolor=blue,citecolor=red,urlcolor=blue]{hyperref}
\usepackage[rightcaption]{sidecap}
\usepackage{array}
\usepackage{multirow}
\usepackage{tabularx}
\usepackage{braket} 
\usepackage{dsfont}
\usepackage{bbold}
\usepackage{etoolbox}

\AtBeginDocument{%
    \newwrite\bibnotes
    \def\bibnotesext{Notes.bib}
    \immediate\openout\bibnotes=\jobname\bibnotesext
    \immediate\write\bibnotes{@CONTROL{REVTEX41Control}}
    \immediate\write\bibnotes{@CONTROL{%
    apsrev41Control,author="08",editor="1",pages="0",title="0",year="1"}}
     \if@filesw
     \immediate\write\@auxout{\string\citation{apsrev41Control}}%
    \fi
}%

\newcommand{\bs}[1]{\boldsymbol{#1}}

\def\half{{\tfrac{1}{2}}} 

\def\Br{{\bs{r}}}
\def\Bk{{\bs{k}}}
\def\Be{{ \bs{e} }}
\def\BR{{\bs{R}}}
\def\BS{{\bs{S}}}
\def\BD{{\bs{D}}}
\def\BPsi{{\bs{\Psi}}}
\def\Ba{{\bs{a}}}
\def\Balpha{{\bs{\alpha}}}

\def\zhat{{ \bs{\hat{z}} }}

\def\muB{{ \mu_{\mbox{\tiny B}} }}

\def\CalO{{\mathcal{O}}}
\def\CalH{{\mathcal{H}}}
\def\CalE{{\mathcal{E}}}

\def\Hsw{{H_{\mbox{\tiny SW}}}}

\def\Tr{{\mbox{Tr}}}

\def\Re{{\mbox{Re}}}

\begin{document}

\title{Chiral Magnonic Edge States in Ferromagnetic Skyrmion Crystals Controlled by Magnetic Fields}

\author{Sebasti{\'a}n A. D{\'i}az}
\affiliation{Department of Physics, University of Basel, Klingelbergstrasse 82, CH-4056 Basel, Switzerland}

\author{Tomoki Hirosawa}
\affiliation{Department of Physics, University of Tokyo, Bunkyo, Tokyo 113-0033, Japan}

\author{Jelena Klinovaja}
\affiliation{Department of Physics, University of Basel, Klingelbergstrasse 82, CH-4056 Basel, Switzerland}

\author{Daniel Loss}
\affiliation{Department of Physics, University of Basel, Klingelbergstrasse 82, CH-4056 Basel, Switzerland}

\date{\today}
	
\begin{abstract}
Achieving control over magnon spin currents in insulating magnets---where dissipation due to Joule heating is highly suppressed---is an active area of research that could lead to energy-efficient spintronics applications. However, magnon spin currents supported by conventional systems with uniform magnetic order have proven hard to control. An alternative approach that relies on topologically protected magnonic edge states of spatially periodic magnetic textures has recently emerged. A prime example of such textures is the ferromagnetic skyrmion crystal which hosts chiral edge states providing a platform for magnon spin currents. Here, we show, for the first time, an external magnetic field can drive a topological phase transition in the spin wave spectrum of a ferromagnetic skyrmion crystal. The topological phase transition is signaled by the closing of a low-energy bulk magnon gap at a critical field. In the topological phase, below the critical field, two topologically protected chiral magnonic edge states lie within this gap, but they unravel in the trivial phase, above the critical field. Remarkably, the topological phase transition involves an inversion of two magnon bands that at the $\Gamma$ point correspond to the breathing and anticlockwise modes of the skyrmions in the crystal. Our findings suggest that an external magnetic field could be used as a knob to switch on and off magnon spin currents carried by topologically protected chiral magnonic edge states.
\end{abstract}

\maketitle


A major obstacle to meet the demand for ever smaller logic and storage devices, based on expectations set by Moore's law, is the pernicious effect of Joule heating \cite{Tu2017}. Two diverging approaches relying on spin degrees of freedom instead of electric charges have come to the rescue. Embracing heating effects and trying to utilize them to produce electronic spin currents is the goal of the field of spin caloritronics \cite{Bauer2012}. Alternatively, avoiding displacing electric charges altogether can be achieved via spin currents carried by magnons in insulating magnets \cite{Meier2003,Kruglyak2010,Demokritov2013,Chumak2015}. While several magnonic analogs of electronic phenomena that could help design the next generation of energy efficient devices have already been identified \cite{Meier2003,Nakata2015b,Nakata2017a,Nakata2017b}, a growing niche for magnon spin current applications that stray from those inspired by conventional electronics has recently emerged \cite{Papp2017,Csaba2017}.

\begin{figure}[t!]
\centering
\includegraphics[width=\columnwidth]{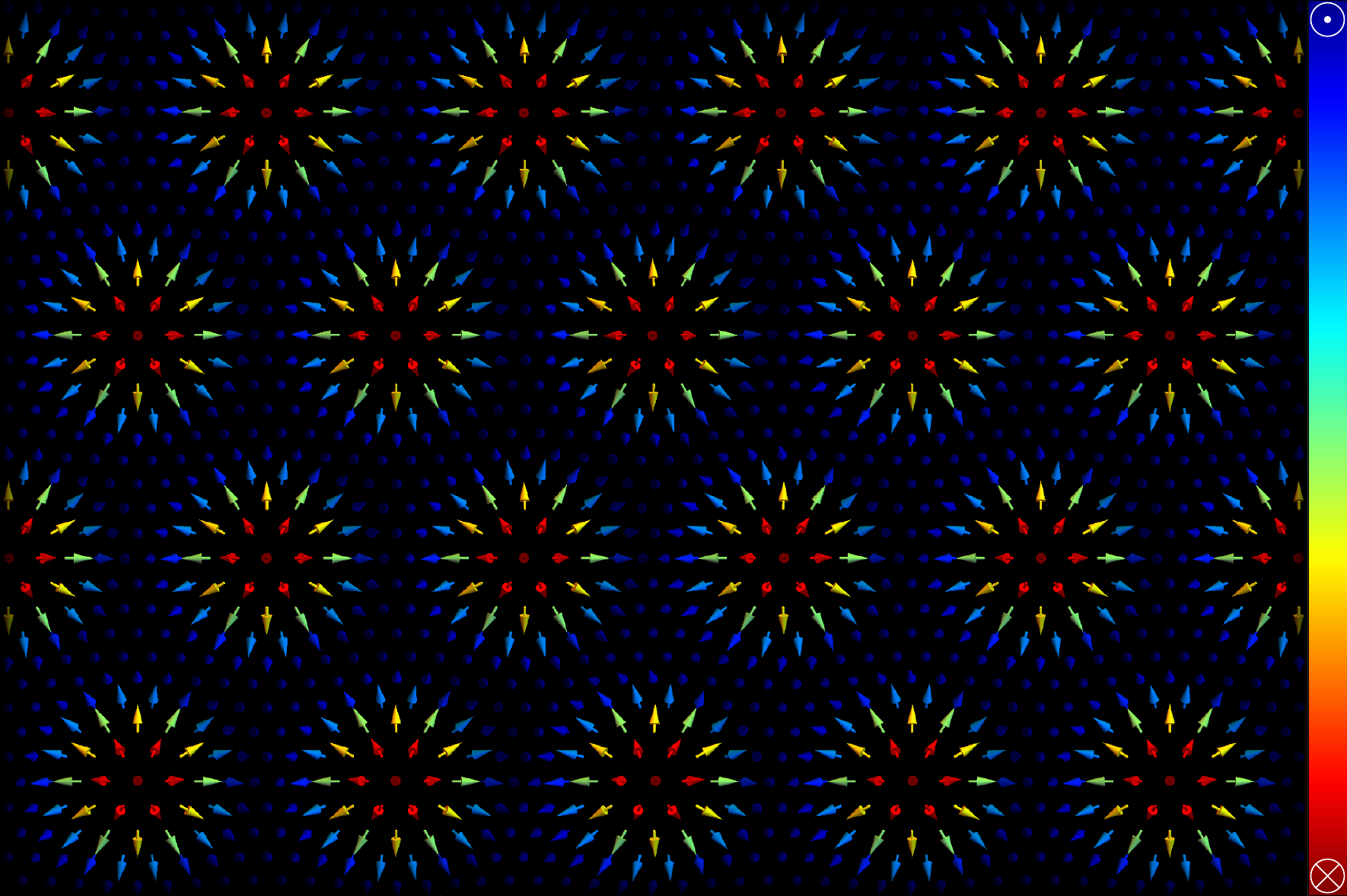}
\caption{{\bf Ferromagnetic skyrmion crystal texture.} Triangular crystal of ferromagnetic N{\'e}el skyrmions obtained from the classical ground-state texture of the spin-lattice Hamiltonian model for $D/J = 1.0$ and $b = 0.8$. The out-of-plane component of the moments is color-coded.}
\label{fig:FMSkX}
\end{figure}

Despite progress on launching and getting a handle on the directionality of magnon spin waves \cite{Schneider2010,Vogt2012,Gieniusz2013,Sadovnikov2015}, reliable control of their propagation and robustness against sample imperfections are still lacking. Realizing such control could be attained in topological magnon insulators \cite{Shindou2013,Zhang2013,Mook2014}, systems with a spatially periodic magnetic texture that support topologically protected magnonic edge states. A promising platform for the nascent field of topological magnonics \cite{Wang2018,Diaz2019,Malki2019} are ferromagnetic skyrmion crystals (FM-SkXs) \cite{Muhlbauer2009,Yu2010}, whose magnetic texure is shown in Fig. \ref{fig:FMSkX}. Although their low energy spin wave modes had been investigated before \cite{Zang2011,Petrova2011,Mochizuki2012,Okamura2013}, only recently they were predicted to host chiral magnonic edge states protected by the nontrivial topology of their bulk magnon bands \cite{Roldan-Molina2016}.

In contrast to electronic topological insulators, ground-state magnetic phases such as FM-SkXs, and consequently the spin wave spectrum they support, depend crucially on applied magnetic fields. Therefore, it is a nontrivial question whether a magnetic field can drive a topological phase transition in the spin wave spectrum without drastically modifying the ground-state texture. Quite remarkably, we discover that there is a window of magnetic field values where such a phase transition between two topologically distinct spin wave spectra of the FM-SkX can be observed. It involves two low-energy bulk magnon bands, which at the $\Gamma$ point correspond to the anticlockwise and breathing modes \cite{Mochizuki2012}. Recent experimental studies in bulk chiral magnets show the energy difference of these modes gets smaller with increasing magnetic field, but it does not vanish upon reaching the boundary with the conical phase \cite{Okamura2013}. On the other hand, our two-dimensional model predicts a critical magnetic field value at which these two bands touch at the $\Gamma$ point, closing a magnon gap and signaling the topological phase transition. Increasing the magnetic field past this critical value reopens the gap, the bands exchange Chern numbers, and band inversion takes place in the vicinity of the gap-closing point. Furthermore, topologically protected chiral magnonic edge states within this gap are only supported in the phase below the critical field. Therefore, an external magnetic field could be used to control the propagation of robust magnon spin currents with a defined directionality along the edge of FM-SkXs in thin films.


\section{Spin Waves in a Ferromagnetic Skyrmion Crystal}

The FM-SkX is obtained from the classical ground-state texture of spin-$S$ operators on a triangular lattice (with lattice constant $a$) governed by the Hamiltonian
\begin{align}\label{eq:SpinH}\nonumber
H &= - \half \sum_{< \Br, \Br' >} \Big(
J_{\Br,\Br'} \BS_\Br\cdot\BS_{\Br'} + \BD_{\Br,\Br'}\cdot\BS_\Br\times\BS_{\Br'} \Big) \\
& \hspace{70pt} + g\muB B \sum_\Br \BS_\Br\cdot\zhat \,,
\end{align}
including nearest neighbor exchange, $J_{\Br,\Br'} = J$, and interfacial Dzyaloshinskii-Moriya (DM) interactions, $\BD_{\Br,\Br'} = D \, \zhat \times (\Br - \Br')/|\Br - \Br'|$, as well as a coupling to an external magnetic field, $B\zhat$, where $g$ is the gyromagnetic ratio and $\muB$ is the Bohr magneton. Focusing on zero temperature, we use Monte Carlo simulated annealing \cite{Evans2014} followed by further time evolution employing a restricted Landau-Lifshitz-Gilbert (LLG) equation \cite{Berkov2005} to obtain the classical ground-state spin texture as a function of the magnetic field. The magnetic moment texture of the FM-SkX is related to the classical ground-state spin texture by $\bs{M}_{\bs{r}} = - \braket{GS | \bs{S}_{\bs{r}} | GS}$, whose direction $\bs{\hat{m}}_{\bs{r}} = \bs{M}_{\bs{r}}/|\bs{M}_{\bs{r}}|$ is shown in Fig. \ref{fig:FMSkX}.

Spin waves are described by rotating the spin quantization axis at each lattice site along the direction of the FM-SkX spin texture so that the rotated spin operators model collective excitations about this classical ground state. Assuming small deviations, we employ a standard transformation mapping the rotated spin operators to Holstein-Primakoff (HP) bosons \cite{Holstein1940}, $a_{\Br}$ and $a^\dag_{\Br}$, to then expand in powers of $1/S$ \cite{Kittel1963}.

The FM-SkX is a triangular crystal of skyrmions. Its magnetic unit cell, a single skyrmion, consists of many spins. This is naturally described by a triangular Bravais lattice, $\BR$, with a basis, $\Br_j$. Therefore, the triangular lattice of spins is partitioned as $\Br = \BR + \Br_j$. The Hamiltonian of free spin waves is then obtained, by taking a lattice Fourier transform, as the $\CalO(S)$ piece in the expansion
\begin{align}\label{eq:Hsw_reciprocal}
\Hsw = \half S \sum_{\Bk;i,j} \BPsi_{\Bk i}^\dag \CalH_{ij}(\Bk) \BPsi_{\Bk j} + \CalE_0 \,.
\end{align}
Here $\BPsi_{\Bk i} = (a_{\Bk i} \,, a_{- \Bk i}^\dag)^T$, $a_{\Bk j} = \tfrac{1}{\sqrt{N}}\sum_\BR e^{-i\Bk\cdot(\BR + \Br_j)}\, a_{\BR j}$, $\CalE_0 = - \half N S\sum_i \Lambda_i$, with $N$ being the total number of magnetic unit cells,
\begin{align}
\CalH_{ij}(\Bk) = 
\begin{pmatrix}
\Omega_{ij}(\Bk) & - \Delta_{ij}(\Bk) \\
- \Delta_{ij}^*(-\Bk) & \Omega^*_{ij}(-\Bk)
\end{pmatrix} \,,
\end{align}
where $\Delta_{ij}(\Bk) =  \half \big[ J_{ij}(\Bk) \, {\Be_i^+ \cdot \Be_j^+} + \BD_{ij}(\Bk) \cdot {\Be_i^+ \times \Be_j^+} \big]$, $\Omega_{ij}(\Bk) = \delta_{ij} \Lambda_i - \half \big[ J_{ij}(\Bk) \, {\Be_i^+ \cdot \Be_j^-} + \BD_{ij}(\Bk) \cdot {\Be_i^+ \times \Be_j^-} \big]$, $\Lambda_i = \sum_j \big[ J_{ij}(\Bk \!\!=\!\! 0) \, {\Be_i^3 \cdot \Be_j^3} + \BD_{ij}(\Bk \!\! =\!\! 0) \cdot {\Be_i^3 \times \Be_j^3} \big] - bJ \, \Be_i^3\cdot\zhat$, with $b = g\muB B/JS$, $J_{ij}(\Bk) = \sum_\BR J_{\BR + \Br_i , \Br_j } e^{- i\Bk \cdot ( \BR + \Br_i - \Br_j )}$ and mutatis mutandis for $\BD_{ij}(\Bk)$. The orthonormal frame $\{ \Be_j^1,\Be_j^2, \Be_j^3 \}$ is defined at each magnetic unit cell site $j$, where $\Be_j^3$ is parallel to the classical ground-state spin texture and $\Be_j^\pm = \Be_j^1 \pm i\,\Be_j^2$.

A paraunitary Bogoliubov transformation, $T_\Bk$, diagonalizes the above spin wave Hamiltonian yielding
\begin{align}
\Hsw = S \sum_{\Bk,\lambda} \CalE_{\Bk \lambda} \big( \alpha_{\Bk \lambda}^\dag \alpha_{\Bk \lambda} + \half \big) + \CalE_0 \,,
\end{align}
where $\alpha_{\Bk \lambda}^\dag$ creates magnons with crystal momentum $\Bk$ at the $\lambda$-th band whose energy is given by $\CalE_{\Bk \lambda}$. Each band can be classified by their Chern number, a topological index defined in reciprocal space, given by an integral over the first Brillouin zone, $C_\lambda = i \int d\Bk \, \epsilon_{\mu\nu} \Tr \{ P_\lambda(\Bk) \, [\partial_{k_\mu} P_\lambda(\Bk)] \, \partial_{k_\nu} P_\lambda(\Bk) \}/2\pi$, with $P_\lambda(\Bk) = T_\Bk \, \Pi_\lambda \, \sigma_z \, T_\Bk^\dag \, \sigma_z$ being a projection operator and $(\Pi_\lambda)_{ij} = \delta_{\lambda i} \delta_{ij}$ \cite{Shindou2013}.

We numerically compute the bulk magnon band structure and the Chern number of the lowest energy bands as a function of external magnetic field. Using magnon coherent states (details in the Supplementary Material) we also obtain the time evolution of the real-space deformation of the FM-SkX for those same bands.


\section{Nearly Flat and Dispersive Bands} 

\begin{figure*}[t!]
\centering
\includegraphics[width=2\columnwidth]{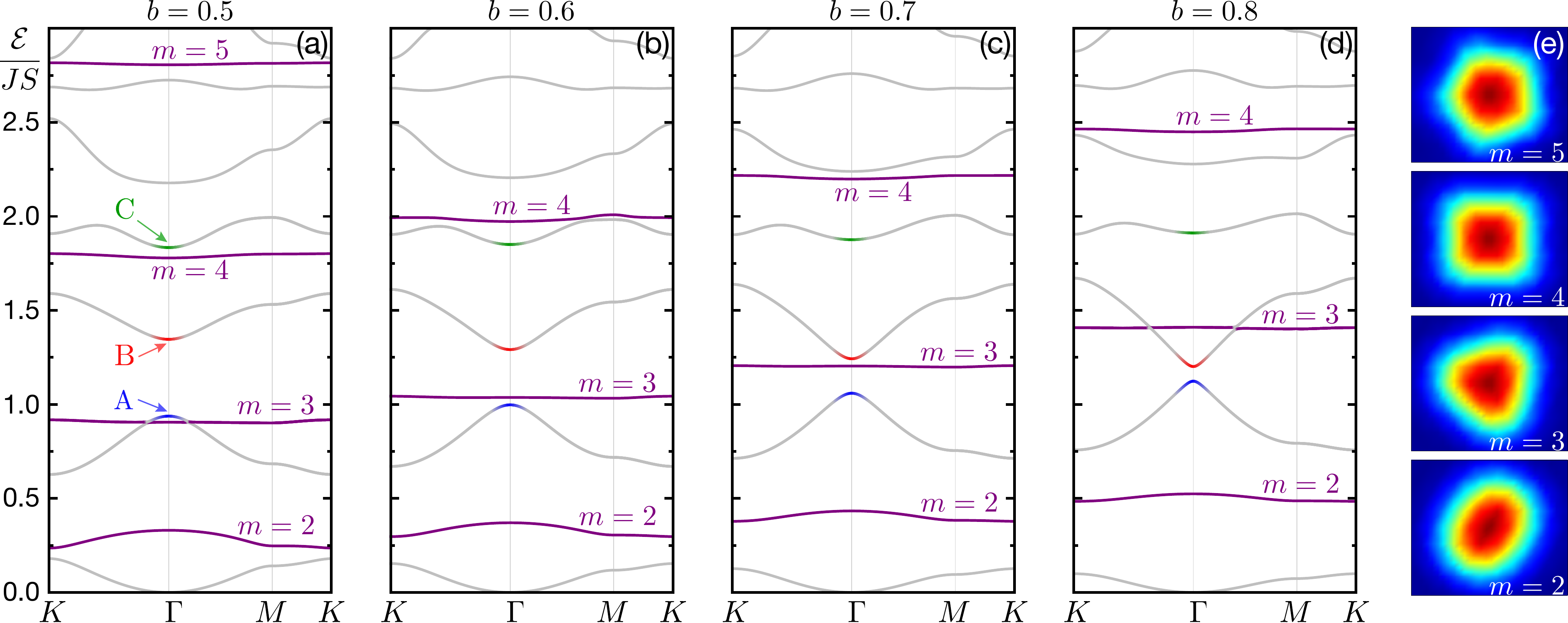}
\caption{{\bf Magnetic field dependence of the magnon spectrum.} (a)-(d) Bulk magnon spectra of the FM-SkX along the first Brillouin zone loop $K$-$\Gamma$-$M$-$K$ for increasing values of the applied magnetic field $b = g\muB B/JS$. Nearly flat (dispersive) bands are depicted in purple (gray). The anticlockwise (A), breathing (B), and clockwise (C) modes are respectively denoted in blue, red, and green. (e) Localized skyrmion distortions corresponding to the nearly flat bands from panels (a)-(d), labeled by their azimuthal number $m$.}
\label{fig:FlatBands}
\end{figure*}

Now we discuss the magnon bands. Figures \ref{fig:FlatBands} (a)-(d) show the lowest energy bulk magnon bands for increasing values of the (rescaled) external magnetic field $b$. The majority of the bands are dispersive, but a few of them appear to be nearly flat. As the magnetic field increases, the nearly flat bands are shifted to higher energies, while the dispersive bands remain almost unaffected. Moreover, overlapping bands belonging to these two different classes hybridize less the lower their energy is. At the $\Gamma$ point, for $\Bk = 0$, time evolution reveals the nearly flat bands correspond to elliptical, triangular, square, and pentagonal distortions of the individual skyrmions in the FM-SkX [see Fig. \ref{fig:FlatBands} (e)]. As time evolves, these localized deformations rotate clockwise, retaining their shape, about the center of each skyrmion and uniformly across the FM-SkX. For $\Bk \neq 0$, the distortions of individual skyrmions are identical up to a phase, $e^{i \Bk\cdot\Br}$, as expected of spatially modulated, propagating spin waves (see Fig. \ref{fig:SWModulation} in the Supplementary Material).

The above modes are expected to be part of the magnon spectrum because they also appear in the spectrum of isolated skyrmions \cite{Lin2014}, and are known to affect their dynamics \cite{Psaroudaki2018}. In analogy to small distortions of a circular domain wall, which can be classified by their azimuthal number $m \in \{ 0, 1, 2, ...\}$, we also label these magnon bands by $m$. Although not as flat as the others, but sharing the same real-space time evolution features, the second lowest energy band corresponds to $m = 2$. Note, however, that it flattens out with increasing magnetic field because, as skyrmions shrink, their overlap with nearest neighbors decreases, hence approaching the isolated skyrmion limit. The lowest energy band corresponds to a localized gyrotropic mode, a remnant of the translational modes ($m = 1$) of the isolated skyrmion \cite{Petrova2011}. Remarkably, the Chern number of all the nearly flat bands is zero.

Among the dispersive modes there are three of special importance: anticlockwise (A), breathing (B), and clockwise (C). They correspond to skyrmion distortions in the vicinity of $\Bk = 0$ [see Figs. \ref{fig:FlatBands} (a)-(d)]. First predicted by LLG simulations \cite{Mochizuki2012}, these three low energy modes can be excited by spatially uniform AC magnetic fields: out-of-plane for the breathing mode and in-plane for the anticlockwise and clockwise modes. They have been observed experimentally in bulk samples of chiral \cite{Onose2012,Okamura2013,Schwarze2015} and polar magnets \cite{Ehlers2016}.

Isolated skyrmions also support a localized breathing mode, with azimuthal number $m = 0$, characterized by axially symmetric distortions of their size, periodically shrinking and expanding. However, the breathing mode present in the FM-SkX magnon spectrum is not localized because of the strong overlap experienced by neighboring skyrmions as they expand.

When a nearly flat band passes through a dispersive one, the spin wave mode associated to each band remains largely the same. Despite their overlap in energy, the bands hardly hybridize with each other. By tracking the Chern number of the lowest energy bands as the magnetic field is increased we observe no exchange of Chern number between nearly flat and dispersive bands. Therefore, these two distinct classes of bands are topologically decoupled.


\section{Topological Phase Transition} 

\begin{figure}[t!]
\centering
\includegraphics[width=\columnwidth]{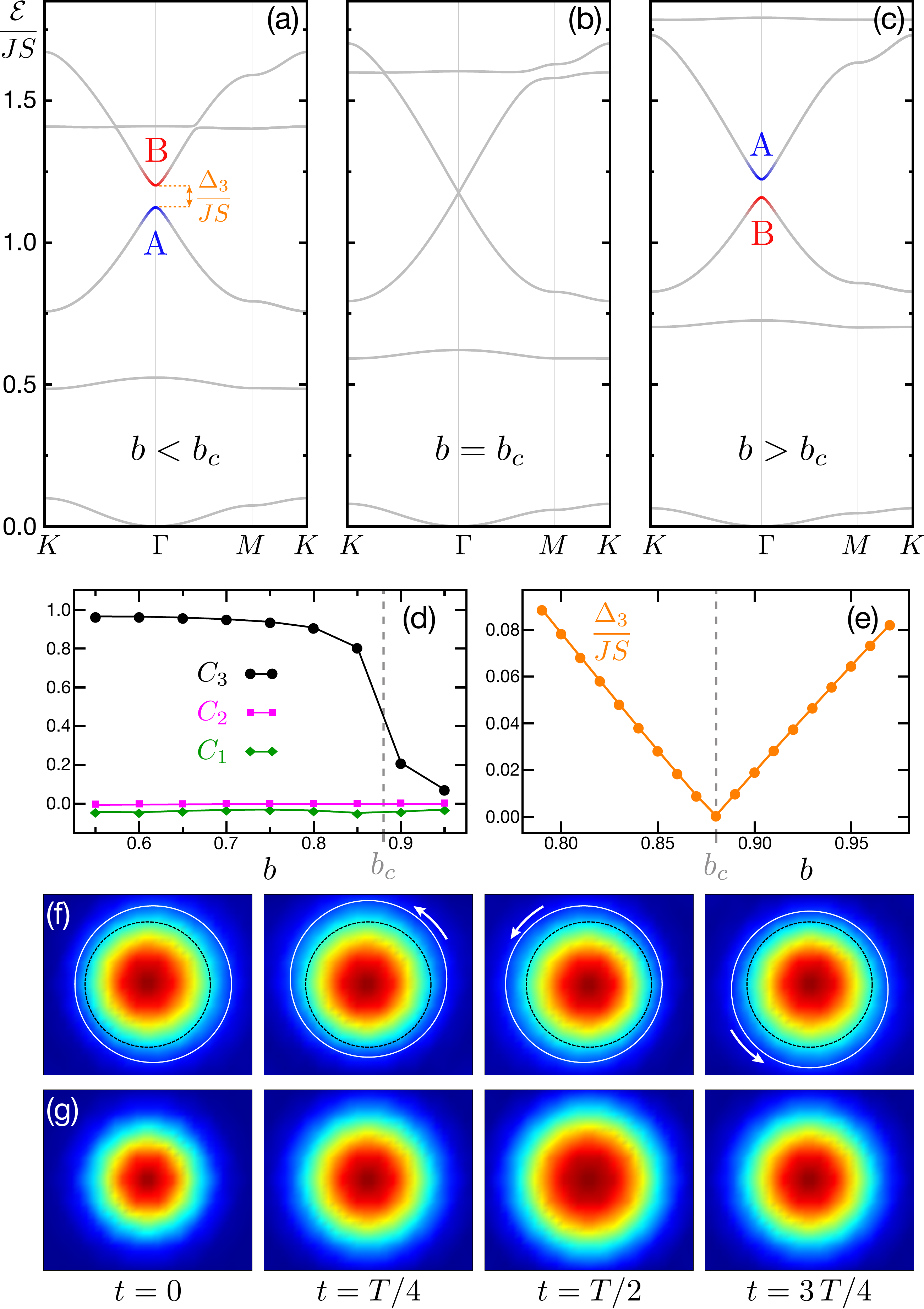}
\caption{{\bf Magnetic field-driven topological phase transition}.
(a)-(c) Bulk magnon spectrum of the FM-SkX in the vicinity of the topological phase transition: (a) for $b=0.80$, (b) for $b = b_c = 0.88$, and (c) for $b=0.95$. The anticlockwise (A) and breathing (B) modes, shown respectively in blue and red, get inverted.
(d) Chern numbers of the three lowest energy bands, $\{C_1,C_2,C_3\}$, as functions of the magnetic field. $C_3$ exhibits a sudden decrease from one to zero as the magnetic field increases past $b_c$, rendering the third band topologically trivial.
(e) Size of the third gap, $\Delta_3$, as a function of the magnetic field, which vanishes at $b_c$ and grows linearly in the vicinity of the topological phase transition.
(f) and (g): Snapshots of the time evolution of the out-of-plane magnetization of the anticlockwise and breathing mode, respectively. Here $T$ denotes the period of both spin wave modes, for simplicity.
}
\label{fig:TopoPhaseTrans}
\end{figure}

Next, we turn to the topological phase transition. As shown in the sequence from Fig. \ref{fig:FlatBands}, the anticlockwise and breathing modes approach each other when the magnetic field increases. By further ramping up the field we observe the third bulk magnon gap closing and then reopening with the two bands touching at the $\Gamma$ point at a critical field $b_c = 0.88$ forming a Dirac-like point [see Fig. \ref{fig:TopoPhaseTrans} (b)]. This gap closing is profoundly different from the others involving nearly flat bands as evidenced by the spin wave modes in the vicinity of the $\Gamma$ point. For magnetic fields smaller than the critical value, $b < b_c$, the third band corresponds to the anticlockwise mode while the fourth band to the breathing mode [see Fig. \ref{fig:TopoPhaseTrans} (a)]. On the other hand, for $b > b_c$, these spin wave modes are inverted [see Fig. \ref{fig:TopoPhaseTrans} (c)]. However, band inversion does not guarantee the occurrence of a topological phase transition \cite{Bansil2016}. 

A better assessment of the nature of the spin wave spectrum changes with magnetic field is given by tracking the Chern numbers. For external magnetic fields larger than the critical value, the third band exhibits a change from $C_3 = 1$ to $C_3 = 0$ [see Fig. \ref{fig:TopoPhaseTrans} (d)], becoming topologically trivial, while the fourth band jumps from $C_4 = 1$ to $C_4 = 2$: a transfer of Chern number takes place. 

Increasing the magnetic field in this range merely shrinks the size of the skyrmions in the FM-SkX without breaking any symmetries. Nevertheless, the bulk magnonic band structure experiences a clear change in its topology. Thus, the system indeed is expected to undergo a topological phase transition at $b_c$. 

\section{Magnonic Edge States and Disorder}

Besides the inversion of the anticlockwise and breathing modes, there is another measurable consequence of the topological phase transition. The bulk-edge correspondence postulates that the number of edge states within the $i$-th bulk gap, $\nu_i$, is related to the Chern numbers of the bulk bands below that gap via $\nu_i = \sum_{j \leq i} C_j$ \cite{Hatsugai1993a,Hatsugai1993b}. Accordingly, since $\nu_3 = 1$, a topologically protected chiral magnonic edge state should be present within the third bulk gap for $b < b_c$, but not for $b_c < b$. However, the predictions of this correspondence still must be checked. As a matter of fact, a stringent requirement of the bulk-edge correspondence is that the finite-sized bulk should consist of an integer number of unit cells \cite{Rhim2018}, which is not necessarily expected to hold for every sample. 

Obtaining a finite system by simply cutting the infinite bulk texture of the FM-SkX arbitrarily is likely to result in a spin configuration that is neither the global nor a local energy minimum, and for which magnons cannot be properly defined. To avoid this issue, we place our system on a generic strip geometry where the edges are a result of sample termination. By recomputing the minimum energy spin configuration of the system in this new geometry, we make sure we obtain more physical bulk and edge textures.

On the new geometry, the FM-SkX reconstructs near the edges and gets slightly distorted (see Fig. \ref{fig:StripEdge} in the Supplementary Material). We choose a sufficiently wide strip to ensure that the system better approximates a large bulk sample with edges. After identifying the new magnetic unit cell we compute the one-dimensional band structure, which typically contains sets of bulk subbands interspersed with ingap edge state bands. Topologically protected edge state bands are expected to extend across the gaps connecting consecutive sets of bulk subbands.

\begin{figure}[t!]
\centering
\includegraphics[width=\columnwidth]{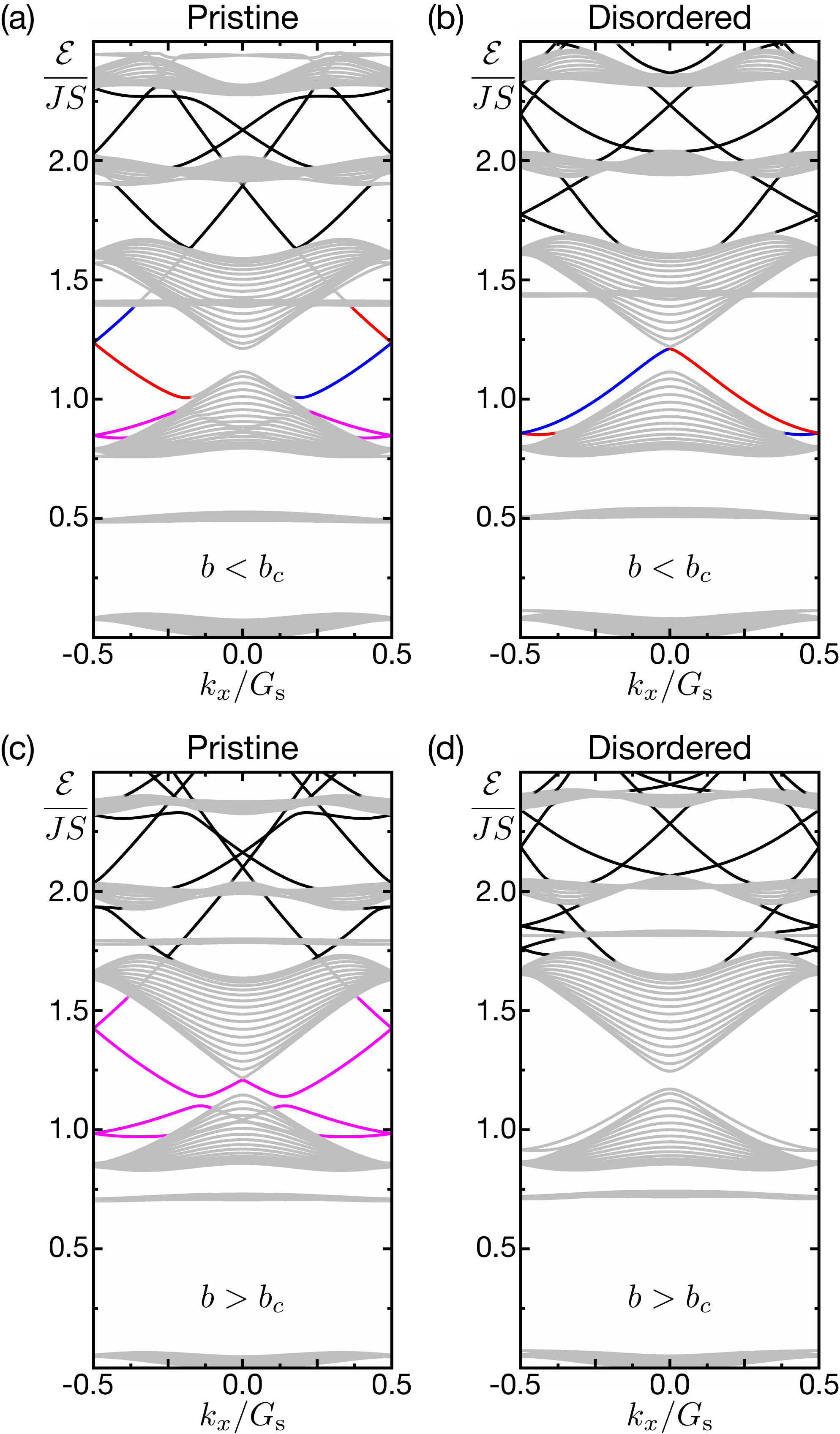}
\caption{{\bf Robustness of magnonic edge states against disorder}.
One-dimensional magnon spectra on an infinite strip: (a) pristine FM-SkX at $b=0.80$, (b) disordered FM-SkX at $b=0.80$, (c) pristine FM-SkX at $b=0.95$, and (d) disordered FM-SkX at $b=0.95$. Bulk subbands are shown in gray and $G_{\rm{s}}$ is the size of the one-dimensional Brillouin zone.
(a) \& (b): For $b < b_c$, the pristine FM-SkX supports topologically protected chiral magnonic edge states within the third gap [right- (left-)moving in blue (red)] which are stable against disorder, while the topologically trivial edge states (purple) are not.
(c) \& (d): For $b > b_c$, the topologically trivial edge states within the third gap (purple) of the pristine FM-SkX, disappear upon introducing disorder.
Edge states within higher energy gaps are indicated in black.
}
\label{fig:EdgeStatesDisorder}
\end{figure}

In Figure \ref{fig:EdgeStatesDisorder} (a) and (c) we show the one-dimensional band structure for $b = 0.80$ and $b = 0.95$, respectively, of the pristine FM-SkX. Consistent with the prediction of the bulk-edge correspondence, while there are no edge states within either the first or second gap, they appear within the third gap for both magnetic field values. However, only in the topological phase, $b < b_c$, edge states extending across the third gap are present (blue and red), while topologically trivial edge states (purple) can be seen in the third gap of both phases. 

To expose the topologically protected edge states while at the same time removing the topologically trivial ones, we introduce a special type of disorder by applying a larger magnetic field along a narrow sliver near the strip edges. The texture is then relaxed to a new, stable magnetic field configuration. As depicted in Figs. \ref{fig:EdgeStatesDisorder} (b) and (d), this model of disorder predominantly changes the edge state bands leaving the bulk subbands mostly invariant. Adding this disorder to the topologically trivial phase completely removes the edge states from the third gap [see Fig. \ref{fig:EdgeStatesDisorder} (d)]. In contrast, only two edge states remain within the third gap in the topological phase [see Fig. \ref{fig:EdgeStatesDisorder} (b)]. Moreover, these edge states are chiral as confirmed by their magnonic wave function. The right-moving (blue) and left-moving (red) edge states are localized at opposite edges of the strip, consistent with a positive winding as predicted by the sign of $\nu_3$ (see the Supplementary Material for details). 

In agreement with the predictions of the bulk-edge correspondence, the above results suggest that the magnonic edge states that remain after introducing disorder are topologically protected. 
Conversely, the edge states that connect a bulk subband set with itself and which unravel with disorder are topologically trivial. We observe two topologically protected edge states in Figs. \ref{fig:EdgeStatesDisorder} (a) and (b) instead of one, as expected from $\nu_3 = 1$, because we use a strip of infinite extension with two spatially separated edges. If we made our strip geometry of finite size, we would observe a single state propagating clockwise along the edge.


\section{Discussion}

The anticlockwise, breathing, and clockwise modes have been measured in bulk samples of semiconducting, metallic, and insulating chiral magnets \cite{Onose2012,Okamura2013,Schwarze2015}. Their reported energy order ($\CalE_A < \CalE_B < \CalE_C$) and their magnetic field dependence are in agreement with our spin wave calculations below the critical field $b_c$. Also similar to our results, the excitation energies of the anticlockwise and breathing modes are observed to approach each other up to the largest field they exist, right at the transition from the FM-SkX to the conical phase. Our two-dimensional model is closer to the physics of thin films that have been suggested to support FM-SkXs at higher fields than in bulk \cite{Roessler2011}. We expect that by extending the stability region of the FM-SkX, the topological phase transition we predict should be observed. 

Throughout this work we took $D/J = 1.0$, while in materials the DM interaction is typically smaller than the exchange. Provided the continuum limit is valid, scaling space and the magnetic field respectively by $aJ/(2D)$ and $3SD^2/(g\muB J)$, maps to a model with equal magnitude of DM interaction and exchange. Thus, for a system with $D/J =1.0$ and one with $D'/J' < 1$, length and magnetic field are related by $l' = l (J'/D')$ and $B' = B(D'/J')^2$. For the realistic parameters $J = 1$ meV, $S = 1$, $D'/J' = 0.05$, and $a = 0.5$ nm, we estimate the skyrmion size as $8 a (J'/D') = 80$ nm and the critical magnetic field as $B'_c = [ b_c JS/(g\muB)] (D'/J')^2 \approx 20$ mT.

State-of-the-art ferromagnetic resonance techniques \cite{Brataas2002,Du2017} could be used to excite either bulk or edge magnonic states. By tuning to an energy window within the third magnon gap, only magnonic edge states should be excited. Their edge localization could be confirmed using Brillouin light scattering \cite{Demokritov2001,Kruglyak2010}. The control of topological phase transitions and the corresponding low-energy topologically protected chiral magnonic edge states by an applied magnetic field, provides a new avenue to exploit robust directional magnon spin currents supported by reconfigurable magnetic textures such as FM-SkXs.

\begin{acknowledgments}

We are grateful to Christina Psaroudaki, Victor Chua, and Alexander Mook for useful discussions. This work was supported by the Swiss National Science Foundation and  NCCR QSIT. This project received funding from the European Union's Horizon 2020 research and innovation program (ERC Starting Grant, grant agreement No 757725).

\end{acknowledgments}

%


\setcounter{figure}{0}
\renewcommand{\thefigure}{S\arabic{figure}}
\renewcommand{\theHfigure}{\thefigure}

\setcounter{equation}{0}
\renewcommand{\theequation}{S\arabic{equation}}
\renewcommand{\theHequation}{\theequation}

\clearpage

\onecolumngrid


\section{Supplementary Information}

\subsection{Bogoliubov Transformation}

To diagonalize the spin wave Hamiltonian $\Hsw$, Eq. \eqref{eq:Hsw_reciprocal} from the main text, the following Bogoliubov transformation must be used
\begin{align}
a_{\Bk j} = U_{\Bk}^{j \lambda} \; \alpha_{\Bk \lambda} + (V_{-\Bk}^{j \lambda})^* \; \alpha_{- \Bk \lambda}^\dag \,, \label{eq:BTrafo}
\end{align}
where the indices $j$ and $\lambda$ run over $\{1, \dots, p \}$ for a magnetic unit cell comprised of $p$ spins. By introducing $\Ba_{\Bk} = (a_{\Bk 1}, \ldots , a_{\Bk p} )^T$ and $\Ba_{\Bk}^\dag = (a_{\Bk 1}^\dag, \ldots , a_{\Bk p}^\dag )^T$, and similarly for $\Balpha_{\Bk}$ and $\Balpha_{\Bk}^\dag$, the Bogoliubov transformation can be written as
\begin{align}
\begin{pmatrix}
\Ba_{\Bk} \\
\Ba_{-\Bk}^\dag
\end{pmatrix}
=
T_\Bk
\begin{pmatrix}
\Balpha_{\Bk} \\
\Balpha_{-\Bk}^\dag
\end{pmatrix} \,,
\end{align}
where 
\begin{align}
T_\Bk = 
\begin{pmatrix}
U_{\Bk} & V_{-\Bk}^* \\
V_{\Bk} & U_{-\Bk}^*
\end{pmatrix} \,,
\end{align}
is a paraunitary matrix that enforces the bosonic algebra on $\alpha_{\Bk \lambda}$ and $\alpha_{\Bk \lambda}^\dag$.


\subsection{Time Evolution of Magnon Modes}

In order to obtain the real-space time evolution of the magnon modes of the ferromagnetic skyrmion crystal, we need to evaluate the expectation value of the spin operators $\BS_{\BR j}$ at each lattice site $\Br = \BR + \Br_j$. The straightforward choice of using the eigenstates of the spin wave Hamiltonian
\begin{align}
\Hsw = S \sum_{\Bk,\lambda} \CalE_{\Bk \lambda} \big( \alpha_{\Bk \lambda}^\dag \alpha_{\Bk \lambda} + \half \big) + \CalE_0 \,,
\end{align}
namely, $\ket{\Bk,\lambda} = \alpha_{\Bk \lambda}^\dag \ket{0}$, has a major disadvantage. As a matter of fact, the expectation value of the spin operator components transverse to the ground-state texture vanish for these eigenstates. While having a fixed number of magnons---hence dubbed number states---these states have an arbitrary phase which makes them unsuitable to describe coherent time evolution of magnon modes. The standard solution to this problem is to resort to magnon coherent states \cite{Zagury1971,Majlis2007}, i.e., eigenstates of the magnon annihilation operators. They are explicitly constructed as  
\begin{align}
\ket{z_{\Bk \lambda}} = e^{ - \frac{1}{2} |z_{\Bk \lambda}|^2} e^{z_{\Bk \lambda} \alpha_{\Bk \lambda}^\dag }\ket{0} \,,
\end{align}
where $\alpha_{\Bk \lambda} \ket{0} = 0$, and they of course satisfy $\alpha_{\Bk \lambda} \ket{z_{\Bk \lambda}} = z_{\Bk \lambda} \ket{z_{\Bk \lambda}}$. Note that $\braket{z_{\Bk \lambda} | \alpha_{\Bk \lambda}^\dag \alpha_{\Bk \lambda} | z_{\Bk \lambda}} = |z_{\Bk \lambda}|^2 = \bar{n}_{\Bk \lambda}$ is the average number of magnons in the $\lambda$-th band with crystal momentum $\Bk$.

\begin{figure}[t!]
\centering
\includegraphics[width=0.8\columnwidth]{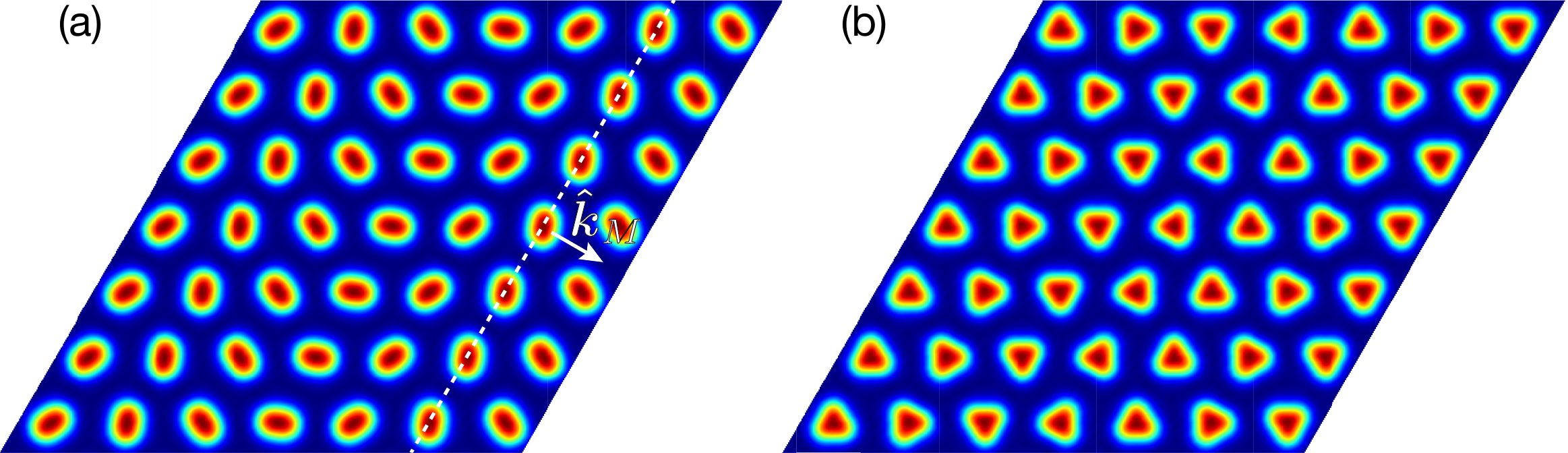}
\caption{{\bf Propagating nearly flat magnon modes.} Snapshots of nearly flat magnon modes of the ferromagnetic skyrmion crystal: (a) elliptical distortion, $m = 2$, and (b) triangular distortion, $m = 3$. For both panels, $\Bk = \frac{1}{2}\Bk_M$, where $\Bk_M$ is a vector in the first Brillouin zone directed from the $\Gamma$ to the $M$ point. The phase fronts are parallel to the dashed white line and simply correspond to locally distorted skyrmions rotated about their centers by the same angle.}
\label{fig:SWModulation}
\end{figure}

To lowest order in Holstein-Primakoff bosons, the spin operators are given by $\BS_{\BR j} = \tfrac{\sqrt{2S}}{2} (\Be_{j}^{-} a_{\BR j} + \Be_{j}^{+} a_{\BR j}^\dag) + S \Be_{j}^{3}$. The expectation value we must now compute reads
\begin{align}
\braket{z_{\Bk \lambda} | \BS_{\BR j}(t) | z_{\Bk \lambda}} = \sqrt{2S} \, \Re \big[ \Be_{j}^{-} \braket{z_{\Bk \lambda} | a_{\BR j}(t) | z_{\Bk \lambda}} \big] + S \Be_{j}^{3} \,.
\end{align}
From the inverse lattice Fourier transform $a_{\BR j} = \frac{1}{\sqrt{N}} \sum_{\Bk} e^{i \Bk \cdot(\BR + \Br_j)} a_{\Bk j}$, where $N$ is the total number of magnetic unit cells, and the Bogoliubov transformation defined in Eq. \eqref{eq:BTrafo}, we notice that computing the time-dependent operator $a_{\BR j}(t)$ reduces to finding $\alpha_{\Bk \lambda}(t)$. Solving the Heisenberg equation
\begin{align}
\frac{d}{dt}\alpha_{\Bk \lambda}(t) = \frac{i}{\hbar} \big[ \Hsw, \alpha_{\Bk \lambda} \big] = - \frac{i}{\hbar} \,S \, \CalE_{\Bk \lambda} \, \alpha_{\Bk \lambda}(t) \,,
\end{align}
we arrive at $\alpha_{\Bk \lambda}(t) = \alpha_{\Bk \lambda} e^{- \frac{i}{\hbar} \,S \, \CalE_{\Bk \lambda}t}$. Employing the above relations, we finally get 
\begin{align}
\braket{z_{\Bk \lambda} | \BS_{\BR j}(t) | z_{\Bk \lambda}} = \sqrt{2S\frac{\bar{n}_{\Bk \lambda}}{N}} \, \Re \left[ \Be_{j}^{-} \left( U_{\Bk}^{j \lambda} e^{i \Phi}  + (V_{\Bk}^{j \lambda})^* e^{- i \Phi} \right) \right] + S \Be_{j}^{3}  \,,
\end{align}
where $\Phi = \Bk \cdot ( \BR + \Br_j ) - \frac{1}{\hbar} S \CalE_{\Bk \lambda} t + \beta_{\Bk \lambda}$, and $\beta_{\Bk \lambda}$ is the phase angle of the magnon coherent state $\ket{z_{\Bk \lambda}}$, i.e., $z_{\Bk \lambda} = \sqrt{\bar{n}_{\Bk \lambda}}\, e^{i \beta_{\Bk \lambda}}$.


\subsection{Strip Geometry}

\subsubsection{Magnetic Texture Near the Edges}

\begin{figure}[h!]
\centering
\includegraphics[width=0.8\columnwidth]{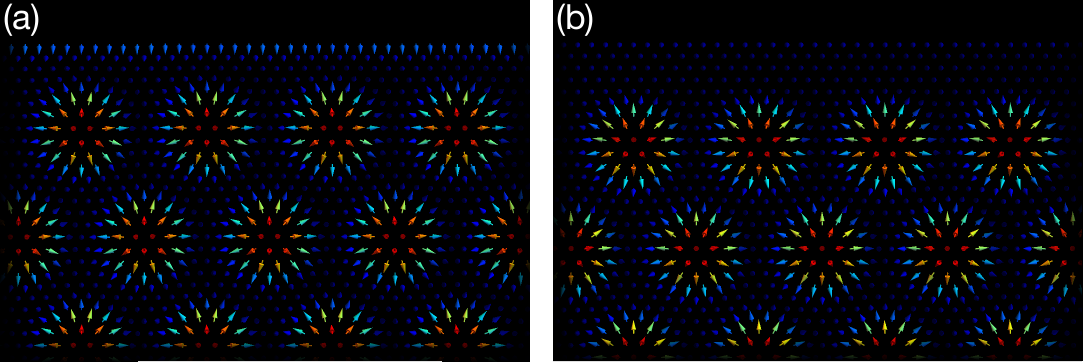}
\caption{{\bf Magnetic texture reconstruction near the strip edges.} (a) Pristine system with a uniform external magnetic field $b=0.80$. (b) Disordered system with an external magnetic field $b=0.80$ everywhere except along the five rows of spins closest to the edge where it was increased to $b=3.0$.}
\label{fig:StripEdge}
\end{figure}

We use an infinitely long strip of finite width to introduce edges. Spins along the edges have less neighboring spins to interact with. Recalculating the minimum energy spin configuration for the strip geometry results in a reconstruction of the magnetic texture near the edges as shown in Fig. \ref{fig:StripEdge} (a). Locally increasing the magnetic field near the edges to model disorder also modifies the magnetic texture, as depicted in  Fig. \ref{fig:StripEdge} (b), effectively pushing the skyrmion crystal further into the bulk of the sample.

\subsubsection{Chiral Magnonic Edge States}

\begin{figure}[t!]
\centering
\includegraphics[width=0.6\columnwidth]{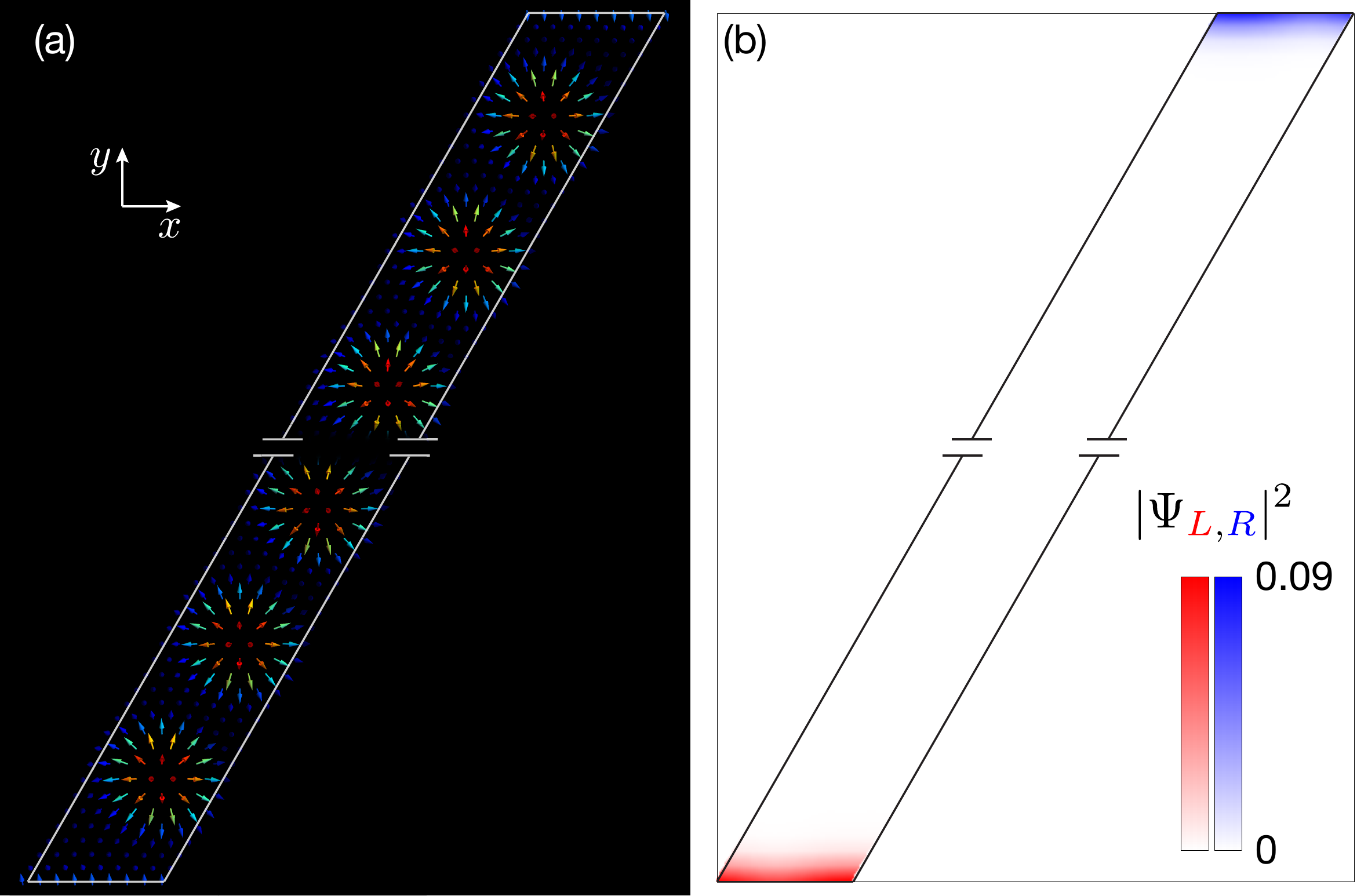}
\caption{{\bf Probability density of  chiral magnonic edge states.} (a) Magnetic unit cell of the pristine ferromagnetic skyrmion crystal in the strip geometry at $b=0.80$. Only the magnetic texture near the top and bottom edges is shown. (b) Probability density $\big| \Psi_{L,R} \big|^2$ of the left-moving (red) and right-moving (blue) magnonic edge states. Both states have the same energy with $k_x = 0.45 G_{\rm{s}}$ for the right-moving state and $k_x = -0.45 G_{\rm{s}}$ for the left-moving state, where $G_{\rm{s}}$ is the size of the one-dimensional Brillouin zone.}
\label{fig:StripWF}
\end{figure}

Chiral magnonic edge states are present within the third magnon gap of the topological phase below the critical field [see Fig. 4 (a) and (b) in the main text]. 
The probability density of the wavefunction of the left- and right-moving edge states is given by
\begin{align}
\big| \Psi_{L/R}(j , k_x) \big|^2 = \big| U_{k_x}^{j R/L} \big|^2 + \big| V_{k_x}^{j R/L} \big|^2 \,,
\end{align}
where $j$ labels the spin sites within the magnetic unit cell of the ferromagnetic skyrmion crystal in the strip geometry [see Fig. \ref{fig:StripWF} (a)] and $k_x$ is the crystal momentum parallel to the strip edges. As depicted in Fig. \ref{fig:StripWF} (b), the probability density  of the right-moving state, $\big| \Psi_{R} \big|^2$, has its weight localized near the top edge, while $\big| \Psi_{L} \big|^2$ is localized near the bottom edge. 

\subsection{Skyrmion Crystal: Spins on Square Lattice}

The ferromagnetic skyrmion crystal from the main text is a magnetic texture of spins that reside on a triangular lattice. We confirmed that our results do not depend on the underlying spin lattice by considering also the case of a ferromagnetic skyrmion crystal supported by a square lattice of spins.

\begin{figure}[b!]
\centering
\includegraphics[width=0.25\columnwidth]{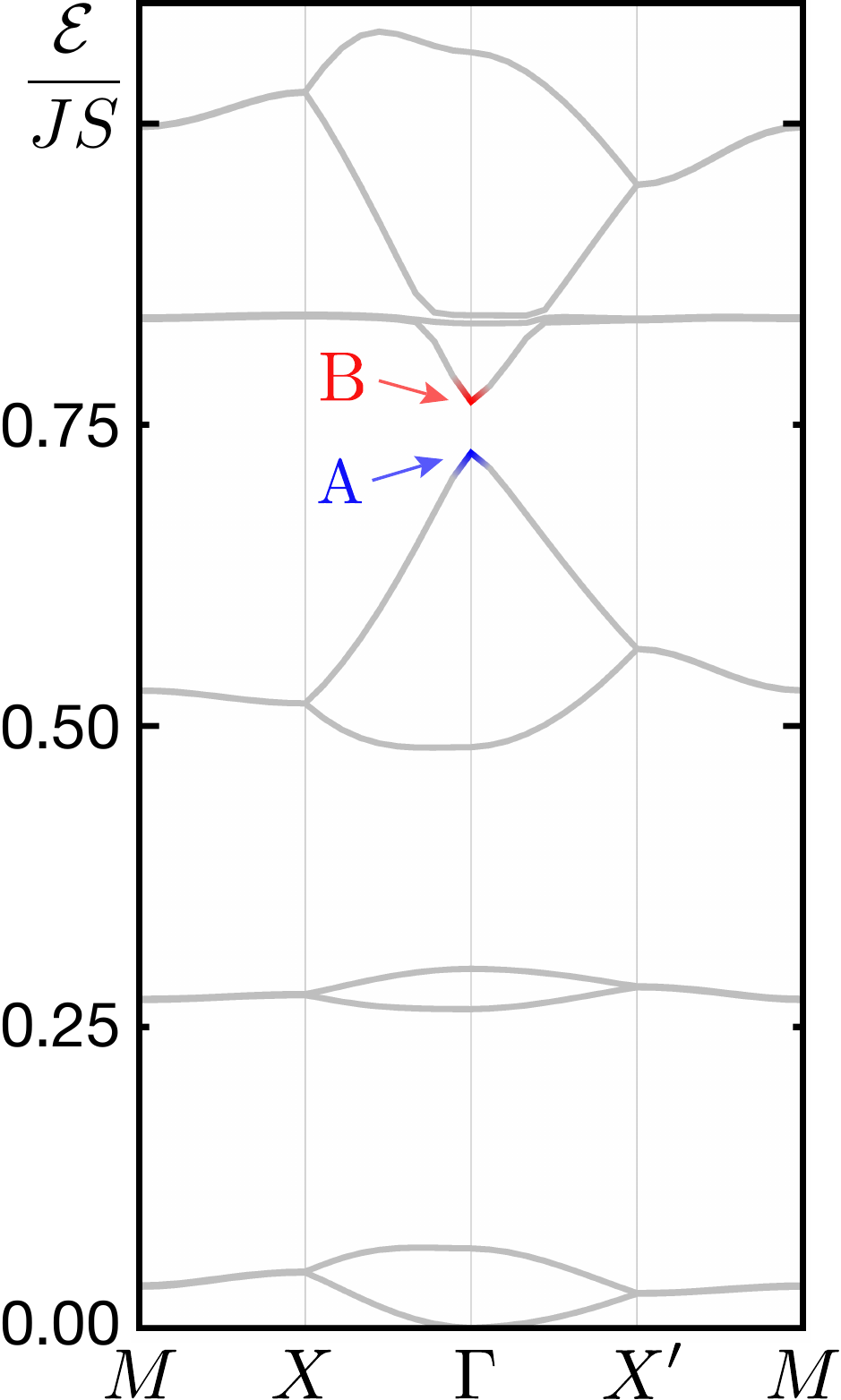}
\caption{{\bf Magnon spectrum} computed for a ferromagnetic skyrmion crystal supported by a square lattice of spins at $b = 0.52$. Bulk magnon bands come in partially degenerate pairs since the magnetic unit cell comprises twice as many spins as the triangular lattice considered in the main text. The anticlockwise (blue) and breathing (red) modes are naturally still present at the $\Gamma$ point.}
\label{fig:SqL_BulkSpectrum}
\end{figure}

The magnetic unit cell is now rectangular and accommodates two skyrmions, which leads to a doubling of the magnon bands. Shown in Fig. \ref{fig:SqL_BulkSpectrum} is the bulk magnon spectrum. Owing to the reduced number of nearest neighbors, the energy scale differs from the triangular lattice case. Nevertheless, a one-to-one correspondence can be established between pairs of partially degenerate magnon bands of the square lattice case and those in the main text. In particular, the nearly flat and dispersive bands---including the anticlockwise, breathing, and clockwise modes at the $\Gamma$ point---are present  in the same energy order.

Similarly, the external magnetic field drives a topological phase transition of the magnon spectrum. At the critical field, numerically estimated to be $b_c = 0.58$, the magnon gap between the anticlockwise and breathing modes closes at the $\Gamma$ point. Furthermore, these two modes invert when the external field is increased past its critical value. This band inversion was independently confirmed from the time evolution of the anticlockwise and breathing modes, obtained using magnon coherent states, as described above, as well as the Landau-Lifshitz-Gilbert equation.

Although the bands in the square lattice case come in partially degenerate pairs, a Chern number for each such pair can still be computed. Just as discussed in the main text, a transfer of Chern number between the bands that invert takes place during the topological phase transition. 

\begin{figure}[t!]
\centering
\includegraphics[width=0.5\columnwidth]{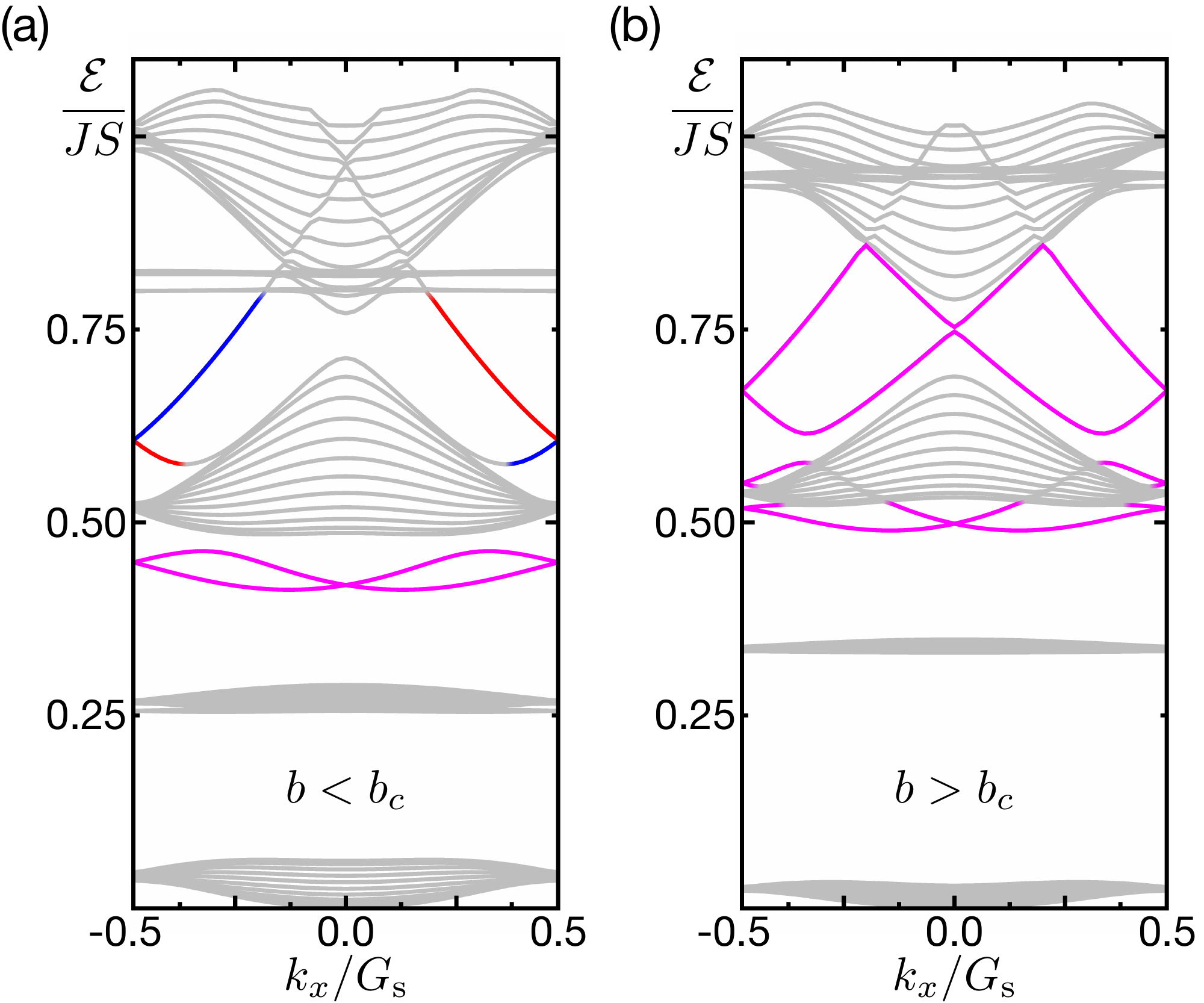}
\caption{{\bf Magnonic edge states.} One-dimensional magnon spectra of a ferromagnetic skyrmion crystal on an infinite strip supported by a square lattice of spins: (a) in the topological phase at $b = 0.52$ and (b) in the trivial phase at $b = 0.60$. Topologically protected chiral magnonic edge states [right-moving (blue) and left-moving (red)] are present within the third gap in the topological phase (a). Topologically trivial edge states (purple) can be seen within the second magnon gap, which could be removed by disorder. Only topologically trivial edge states (purple) are present in the energy window of the trivial phase shown in (b).}
\label{fig:SqL_StripSpectra}
\end{figure}

Below the critical field, in the topological phase, chiral magnonic edge states are present within the third gap [see Fig. \ref{fig:SqL_StripSpectra} (a)]. On the other hand, only topologically trivial states may be found within this gap [see Fig. \ref{fig:SqL_StripSpectra} (b)] above the critical field.

\begin{thebibliography}{47}%
\makeatletter
\providecommand \@ifxundefined [1]{%
 \@ifx{#1\undefined}
}%
\providecommand \@ifnum [1]{%
 \ifnum #1\expandafter \@firstoftwo
 \else \expandafter \@secondoftwo
 \fi
}%
\providecommand \@ifx [1]{%
 \ifx #1\expandafter \@firstoftwo
 \else \expandafter \@secondoftwo
 \fi
}%
\providecommand \natexlab [1]{#1}%
\providecommand \enquote  [1]{``#1''}%
\providecommand \bibnamefont  [1]{#1}%
\providecommand \bibfnamefont [1]{#1}%
\providecommand \citenamefont [1]{#1}%
\providecommand \href@noop [0]{\@secondoftwo}%
\providecommand \href [0]{\begingroup \@sanitize@url \@href}%
\providecommand \@href[1]{\@@startlink{#1}\@@href}%
\providecommand \@@href[1]{\endgroup#1\@@endlink}%
\providecommand \@sanitize@url [0]{\catcode `\\12\catcode `\$12\catcode
  `\&12\catcode `\#12\catcode `\^12\catcode `\_12\catcode `\%12\relax}%
\providecommand \@@startlink[1]{}%
\providecommand \@@endlink[0]{}%
\providecommand \url  [0]{\begingroup\@sanitize@url \@url }%
\providecommand \@url [1]{\endgroup\@href {#1}{\urlprefix }}%
\providecommand \urlprefix  [0]{URL }%
\providecommand \Eprint [0]{\href }%
\renewcommand{\doibase}[1]{https://dx.doi.org/\ifdefempty{#1}{}{#1}}%
\providecommand \selectlanguage [0]{\@gobble}%
\providecommand \bibinfo  [0]{\@secondoftwo}%
\providecommand \bibfield  [0]{\@secondoftwo}%
\providecommand \translation [1]{[#1]}%
\providecommand \BibitemOpen [0]{}%
\providecommand \bibitemStop [0]{}%
\providecommand \bibitemNoStop [0]{.\EOS\space}%
\providecommand \EOS [0]{\spacefactor3000\relax}%
\providecommand \BibitemShut  [1]{\csname bibitem#1\endcsname}%
\let\auto@bib@innerbib\@empty
\bibitem [{\citenamefont {Tu}\ \emph {et~al.}(2017)\citenamefont {Tu},
  \citenamefont {Liu},\ and\ \citenamefont {Li}}]{Tu2017}%
  \BibitemOpen
  \bibfield  {author} {\bibinfo {author} {\bibfnamefont {K.~N.}\ \bibnamefont
  {Tu}}, \bibinfo {author} {\bibfnamefont {Y.}~\bibnamefont {Liu}}, \ and\
  \bibinfo {author} {\bibfnamefont {M.}~\bibnamefont {Li}},\ }\bibfield
  {title} {\enquote {\bibinfo {title} {Effect of joule heating and current
  crowding on electromigration in mobile technology},}\ }\href {\doibase
  10.1063/1.4974168} {\bibfield  {journal} {\bibinfo  {journal} {Applied
  Physics Reviews}\ }\textbf {\bibinfo {volume} {4}},\ \bibinfo {pages}
  {011101} (\bibinfo {year} {2017})}\BibitemShut {NoStop}%
\bibitem [{\citenamefont {Bauer}\ \emph {et~al.}(2012)\citenamefont {Bauer},
  \citenamefont {Saitoh},\ and\ \citenamefont {van Wees}}]{Bauer2012}%
  \BibitemOpen
  \bibfield  {author} {\bibinfo {author} {\bibfnamefont {G.~E.~W.}\
  \bibnamefont {Bauer}}, \bibinfo {author} {\bibfnamefont {E.}~\bibnamefont
  {Saitoh}}, \ and\ \bibinfo {author} {\bibfnamefont {B.~J.}\ \bibnamefont {van
  Wees}},\ }\bibfield  {title} {\enquote {\bibinfo {title} {Spin
  caloritronics},}\ }\href {\doibase 10.1038/nmat3301} {\bibfield  {journal}
  {\bibinfo  {journal} {Nature Materials}\ }\textbf {\bibinfo {volume} {11}},\
  \bibinfo {pages} {391} (\bibinfo {year} {2012})}\BibitemShut {NoStop}%
\bibitem [{\citenamefont {Meier}\ and\ \citenamefont {Loss}(2003)}]{Meier2003}%
  \BibitemOpen
  \bibfield  {author} {\bibinfo {author} {\bibfnamefont {F.}~\bibnamefont
  {Meier}}\ and\ \bibinfo {author} {\bibfnamefont {D.}~\bibnamefont {Loss}},\
  }\bibfield  {title} {\enquote {\bibinfo {title} {Magnetization transport and
  quantized spin conductance},}\ }\href {\doibase
  10.1103/PhysRevLett.90.167204} {\bibfield  {journal} {\bibinfo  {journal}
  {Phys. Rev. Lett.}\ }\textbf {\bibinfo {volume} {90}},\ \bibinfo {pages}
  {167204} (\bibinfo {year} {2003})}\BibitemShut {NoStop}%
\bibitem [{\citenamefont {Kruglyak}\ \emph {et~al.}(2010)\citenamefont
  {Kruglyak}, \citenamefont {Demokritov},\ and\ \citenamefont
  {Grundler}}]{Kruglyak2010}%
  \BibitemOpen
  \bibfield  {author} {\bibinfo {author} {\bibfnamefont {V.~V.}\ \bibnamefont
  {Kruglyak}}, \bibinfo {author} {\bibfnamefont {S.~O.}\ \bibnamefont
  {Demokritov}}, \ and\ \bibinfo {author} {\bibfnamefont {D.}~\bibnamefont
  {Grundler}},\ }\bibfield  {title} {\enquote {\bibinfo {title} {Magnonics},}\
  }\href {http://stacks.iop.org/0022-3727/43/i=26/a=264001} {\bibfield
  {journal} {\bibinfo  {journal} {J. Phys. D: Appl. Phys.}\ }\textbf {\bibinfo
  {volume} {43}},\ \bibinfo {pages} {264001} (\bibinfo {year}
  {2010})}\BibitemShut {NoStop}%
\bibitem [{\citenamefont {Demokritov}\ and\ \citenamefont
  {Slavin}(2013)}]{Demokritov2013}%
  \BibitemOpen
  \bibfield  {author} {\bibinfo {author} {\bibfnamefont {S.}~\bibnamefont
  {Demokritov}}\ and\ \bibinfo {author} {\bibfnamefont {A.}~\bibnamefont
  {Slavin}},\ }\href@noop {} {\emph {\bibinfo {title} {Magnonics: From
  Fundamentals to Applications}}}\ (\bibinfo  {publisher} {Springer-Verlag
  Berlin Heidelberg},\ \bibinfo {year} {2013})\BibitemShut {NoStop}%
\bibitem [{\citenamefont {Chumak}\ \emph {et~al.}(2015)\citenamefont {Chumak},
  \citenamefont {Vasyuchka}, \citenamefont {Serga},\ and\ \citenamefont
  {Hillebrands}}]{Chumak2015}%
  \BibitemOpen
  \bibfield  {author} {\bibinfo {author} {\bibfnamefont {A.~V.}\ \bibnamefont
  {Chumak}}, \bibinfo {author} {\bibfnamefont {V.~I.}\ \bibnamefont
  {Vasyuchka}}, \bibinfo {author} {\bibfnamefont {A.~A.}\ \bibnamefont
  {Serga}}, \ and\ \bibinfo {author} {\bibfnamefont {B.}~\bibnamefont
  {Hillebrands}},\ }\bibfield  {title} {\enquote {\bibinfo {title} {Magnon
  spintronics},}\ }\href {http://dx.doi.org/10.1038/nphys3347} {\bibfield
  {journal} {\bibinfo  {journal} {Nat. Phys.}\ }\textbf {\bibinfo {volume}
  {11}},\ \bibinfo {pages} {453} (\bibinfo {year} {2015})}\BibitemShut
  {NoStop}%
\bibitem [{\citenamefont {Nakata}\ \emph {et~al.}(2015)\citenamefont {Nakata},
  \citenamefont {Simon},\ and\ \citenamefont {Loss}}]{Nakata2015b}%
  \BibitemOpen
  \bibfield  {author} {\bibinfo {author} {\bibfnamefont {K.}~\bibnamefont
  {Nakata}}, \bibinfo {author} {\bibfnamefont {P.}~\bibnamefont {Simon}}, \
  and\ \bibinfo {author} {\bibfnamefont {D.}~\bibnamefont {Loss}},\ }\bibfield
  {title} {\enquote {\bibinfo {title} {Wiedemann-franz law for magnon
  transport},}\ }\href {\doibase 10.1103/PhysRevB.92.134425} {\bibfield
  {journal} {\bibinfo  {journal} {Phys. Rev. B}\ }\textbf {\bibinfo {volume}
  {92}},\ \bibinfo {pages} {134425} (\bibinfo {year} {2015})}\BibitemShut
  {NoStop}%
\bibitem [{\citenamefont {Nakata}\ \emph
  {et~al.}(2017{\natexlab{a}})\citenamefont {Nakata}, \citenamefont {Simon},\
  and\ \citenamefont {Loss}}]{Nakata2017a}%
  \BibitemOpen
  \bibfield  {author} {\bibinfo {author} {\bibfnamefont {K.}~\bibnamefont
  {Nakata}}, \bibinfo {author} {\bibfnamefont {P.}~\bibnamefont {Simon}}, \
  and\ \bibinfo {author} {\bibfnamefont {D.}~\bibnamefont {Loss}},\ }\bibfield
  {title} {\enquote {\bibinfo {title} {Spin currents and magnon dynamics in
  insulating magnets},}\ }\href
  {http://stacks.iop.org/0022-3727/50/i=11/a=114004} {\bibfield  {journal}
  {\bibinfo  {journal} {J. Phys. D: Appl. Phys.}\ }\textbf {\bibinfo {volume}
  {50}},\ \bibinfo {pages} {114004} (\bibinfo {year}
  {2017}{\natexlab{a}})}\BibitemShut {NoStop}%
\bibitem [{\citenamefont {Nakata}\ \emph
  {et~al.}(2017{\natexlab{b}})\citenamefont {Nakata}, \citenamefont
  {Klinovaja},\ and\ \citenamefont {Loss}}]{Nakata2017b}%
  \BibitemOpen
  \bibfield  {author} {\bibinfo {author} {\bibfnamefont {K.}~\bibnamefont
  {Nakata}}, \bibinfo {author} {\bibfnamefont {J.}~\bibnamefont {Klinovaja}}, \
  and\ \bibinfo {author} {\bibfnamefont {D.}~\bibnamefont {Loss}},\ }\bibfield
  {title} {\enquote {\bibinfo {title} {Magnonic quantum hall effect and
  wiedemann-franz law},}\ }\href {\doibase 10.1103/PhysRevB.95.125429}
  {\bibfield  {journal} {\bibinfo  {journal} {Phys. Rev. B}\ }\textbf {\bibinfo
  {volume} {95}},\ \bibinfo {pages} {125429} (\bibinfo {year}
  {2017}{\natexlab{b}})}\BibitemShut {NoStop}%
\bibitem [{\citenamefont {Papp}\ \emph {et~al.}(2017)\citenamefont {Papp},
  \citenamefont {Porod}, \citenamefont {Csurgay},\ and\ \citenamefont
  {Csaba}}]{Papp2017}%
  \BibitemOpen
  \bibfield  {author} {\bibinfo {author} {\bibfnamefont {{\'A}.}~\bibnamefont
  {Papp}}, \bibinfo {author} {\bibfnamefont {W.}~\bibnamefont {Porod}},
  \bibinfo {author} {\bibfnamefont {{\'A}.~I.}\ \bibnamefont {Csurgay}}, \ and\
  \bibinfo {author} {\bibfnamefont {G.}~\bibnamefont {Csaba}},\ }\bibfield
  {title} {\enquote {\bibinfo {title} {Nanoscale spectrum analyzer based on
  spin-wave interference},}\ }\href {\doibase 10.1038/s41598-017-09485-7}
  {\bibfield  {journal} {\bibinfo  {journal} {Scientific Reports}\ }\textbf
  {\bibinfo {volume} {7}},\ \bibinfo {pages} {9245} (\bibinfo {year}
  {2017})}\BibitemShut {NoStop}%
\bibitem [{\citenamefont {Csaba}\ \emph {et~al.}(2017)\citenamefont {Csaba},
  \citenamefont {Papp},\ and\ \citenamefont {Porod}}]{Csaba2017}%
  \BibitemOpen
  \bibfield  {author} {\bibinfo {author} {\bibfnamefont {G.}~\bibnamefont
  {Csaba}}, \bibinfo {author} {\bibfnamefont {{\'A}.}~\bibnamefont {Papp}}, \
  and\ \bibinfo {author} {\bibfnamefont {W.}~\bibnamefont {Porod}},\ }\bibfield
   {title} {\enquote {\bibinfo {title} {Perspectives of using spin waves for
  computing and signal processing},}\ }\href {\doibase
  https://doi.org/10.1016/j.physleta.2017.02.042} {\bibfield  {journal}
  {\bibinfo  {journal} {Physics Letters A}\ }\textbf {\bibinfo {volume}
  {381}},\ \bibinfo {pages} {1471 } (\bibinfo {year} {2017})}\BibitemShut
  {NoStop}%
\bibitem [{\citenamefont {Schneider}\ \emph {et~al.}(2010)\citenamefont
  {Schneider}, \citenamefont {Serga}, \citenamefont {Chumak}, \citenamefont
  {Sandweg}, \citenamefont {Trudel}, \citenamefont {Wolff}, \citenamefont
  {Kostylev}, \citenamefont {Tiberkevich}, \citenamefont {Slavin},\ and\
  \citenamefont {Hillebrands}}]{Schneider2010}%
  \BibitemOpen
  \bibfield  {author} {\bibinfo {author} {\bibfnamefont {T.}~\bibnamefont
  {Schneider}}, \bibinfo {author} {\bibfnamefont {A.~A.}\ \bibnamefont
  {Serga}}, \bibinfo {author} {\bibfnamefont {A.~V.}\ \bibnamefont {Chumak}},
  \bibinfo {author} {\bibfnamefont {C.~W.}\ \bibnamefont {Sandweg}}, \bibinfo
  {author} {\bibfnamefont {S.}~\bibnamefont {Trudel}}, \bibinfo {author}
  {\bibfnamefont {S.}~\bibnamefont {Wolff}}, \bibinfo {author} {\bibfnamefont
  {M.~P.}\ \bibnamefont {Kostylev}}, \bibinfo {author} {\bibfnamefont {V.~S.}\
  \bibnamefont {Tiberkevich}}, \bibinfo {author} {\bibfnamefont {A.~N.}\
  \bibnamefont {Slavin}}, \ and\ \bibinfo {author} {\bibfnamefont
  {B.}~\bibnamefont {Hillebrands}},\ }\bibfield  {title} {\enquote {\bibinfo
  {title} {Nondiffractive subwavelength wave beams in a medium with externally
  controlled anisotropy},}\ }\href {\doibase 10.1103/PhysRevLett.104.197203}
  {\bibfield  {journal} {\bibinfo  {journal} {Phys. Rev. Lett.}\ }\textbf
  {\bibinfo {volume} {104}},\ \bibinfo {pages} {197203} (\bibinfo {year}
  {2010})}\BibitemShut {NoStop}%
\bibitem [{\citenamefont {Vogt}\ \emph {et~al.}(2012)\citenamefont {Vogt},
  \citenamefont {Schultheiss}, \citenamefont {Jain}, \citenamefont {Pearson},
  \citenamefont {Hoffmann}, \citenamefont {Bader},\ and\ \citenamefont
  {Hillebrands}}]{Vogt2012}%
  \BibitemOpen
  \bibfield  {author} {\bibinfo {author} {\bibfnamefont {K.}~\bibnamefont
  {Vogt}}, \bibinfo {author} {\bibfnamefont {H.}~\bibnamefont {Schultheiss}},
  \bibinfo {author} {\bibfnamefont {S.}~\bibnamefont {Jain}}, \bibinfo {author}
  {\bibfnamefont {J.~E.}\ \bibnamefont {Pearson}}, \bibinfo {author}
  {\bibfnamefont {A.}~\bibnamefont {Hoffmann}}, \bibinfo {author}
  {\bibfnamefont {S.~D.}\ \bibnamefont {Bader}}, \ and\ \bibinfo {author}
  {\bibfnamefont {B.}~\bibnamefont {Hillebrands}},\ }\bibfield  {title}
  {\enquote {\bibinfo {title} {Spin waves turning a corner},}\ }\href {\doibase
  10.1063/1.4738887} {\bibfield  {journal} {\bibinfo  {journal} {Applied
  Physics Letters}\ }\textbf {\bibinfo {volume} {101}},\ \bibinfo {pages}
  {042410} (\bibinfo {year} {2012})}\BibitemShut {NoStop}%
\bibitem [{\citenamefont {Gieniusz}\ \emph {et~al.}(2013)\citenamefont
  {Gieniusz}, \citenamefont {Ulrichs}, \citenamefont {Bessonov}, \citenamefont
  {Guzowska}, \citenamefont {Stognii},\ and\ \citenamefont
  {Maziewski}}]{Gieniusz2013}%
  \BibitemOpen
  \bibfield  {author} {\bibinfo {author} {\bibfnamefont {R.}~\bibnamefont
  {Gieniusz}}, \bibinfo {author} {\bibfnamefont {H.}~\bibnamefont {Ulrichs}},
  \bibinfo {author} {\bibfnamefont {V.~D.}\ \bibnamefont {Bessonov}}, \bibinfo
  {author} {\bibfnamefont {U.}~\bibnamefont {Guzowska}}, \bibinfo {author}
  {\bibfnamefont {A.~I.}\ \bibnamefont {Stognii}}, \ and\ \bibinfo {author}
  {\bibfnamefont {A.}~\bibnamefont {Maziewski}},\ }\bibfield  {title} {\enquote
  {\bibinfo {title} {Single antidot as a passive way to create caustic
  spin-wave beams in yttrium iron garnet films},}\ }\href {\doibase
  10.1063/1.4795293} {\bibfield  {journal} {\bibinfo  {journal} {Applied
  Physics Letters}\ }\textbf {\bibinfo {volume} {102}},\ \bibinfo {pages}
  {102409} (\bibinfo {year} {2013})}\BibitemShut {NoStop}%
\bibitem [{\citenamefont {Sadovnikov}\ \emph {et~al.}(2015)\citenamefont
  {Sadovnikov}, \citenamefont {Davies}, \citenamefont {Grishin}, \citenamefont
  {Kruglyak}, \citenamefont {Romanenko}, \citenamefont {Sharaevskii},\ and\
  \citenamefont {Nikitov}}]{Sadovnikov2015}%
  \BibitemOpen
  \bibfield  {author} {\bibinfo {author} {\bibfnamefont {A.~V.}\ \bibnamefont
  {Sadovnikov}}, \bibinfo {author} {\bibfnamefont {C.~S.}\ \bibnamefont
  {Davies}}, \bibinfo {author} {\bibfnamefont {S.~V.}\ \bibnamefont {Grishin}},
  \bibinfo {author} {\bibfnamefont {V.~V.}\ \bibnamefont {Kruglyak}}, \bibinfo
  {author} {\bibfnamefont {D.~V.}\ \bibnamefont {Romanenko}}, \bibinfo {author}
  {\bibfnamefont {Y.~P.}\ \bibnamefont {Sharaevskii}}, \ and\ \bibinfo {author}
  {\bibfnamefont {S.~A.}\ \bibnamefont {Nikitov}},\ }\bibfield  {title}
  {\enquote {\bibinfo {title} {Magnonic beam splitter: The building block of
  parallel magnonic circuitry},}\ }\href {\doibase 10.1063/1.4921206}
  {\bibfield  {journal} {\bibinfo  {journal} {Applied Physics Letters}\
  }\textbf {\bibinfo {volume} {106}},\ \bibinfo {pages} {192406} (\bibinfo
  {year} {2015})}\BibitemShut {NoStop}%
\bibitem [{\citenamefont {Shindou}\ \emph {et~al.}(2013)\citenamefont
  {Shindou}, \citenamefont {Matsumoto}, \citenamefont {Murakami},\ and\
  \citenamefont {Ohe}}]{Shindou2013}%
  \BibitemOpen
  \bibfield  {author} {\bibinfo {author} {\bibfnamefont {R.}~\bibnamefont
  {Shindou}}, \bibinfo {author} {\bibfnamefont {R.}~\bibnamefont {Matsumoto}},
  \bibinfo {author} {\bibfnamefont {S.}~\bibnamefont {Murakami}}, \ and\
  \bibinfo {author} {\bibfnamefont {J.-i.}\ \bibnamefont {Ohe}},\ }\bibfield
  {title} {\enquote {\bibinfo {title} {Topological chiral magnonic edge mode in
  a magnonic crystal},}\ }\href {\doibase 10.1103/PhysRevB.87.174427}
  {\bibfield  {journal} {\bibinfo  {journal} {Phys. Rev. B}\ }\textbf {\bibinfo
  {volume} {87}},\ \bibinfo {pages} {174427} (\bibinfo {year}
  {2013})}\BibitemShut {NoStop}%
\bibitem [{\citenamefont {Zhang}\ \emph {et~al.}(2013)\citenamefont {Zhang},
  \citenamefont {Ren}, \citenamefont {Wang},\ and\ \citenamefont
  {Li}}]{Zhang2013}%
  \BibitemOpen
  \bibfield  {author} {\bibinfo {author} {\bibfnamefont {L.}~\bibnamefont
  {Zhang}}, \bibinfo {author} {\bibfnamefont {J.}~\bibnamefont {Ren}}, \bibinfo
  {author} {\bibfnamefont {J.-S.}\ \bibnamefont {Wang}}, \ and\ \bibinfo
  {author} {\bibfnamefont {B.}~\bibnamefont {Li}},\ }\bibfield  {title}
  {\enquote {\bibinfo {title} {Topological magnon insulator in insulating
  ferromagnet},}\ }\href {\doibase 10.1103/PhysRevB.87.144101} {\bibfield
  {journal} {\bibinfo  {journal} {Phys. Rev. B}\ }\textbf {\bibinfo {volume}
  {87}},\ \bibinfo {pages} {144101} (\bibinfo {year} {2013})}\BibitemShut
  {NoStop}%
\bibitem [{\citenamefont {Mook}\ \emph {et~al.}(2014)\citenamefont {Mook},
  \citenamefont {Henk},\ and\ \citenamefont {Mertig}}]{Mook2014}%
  \BibitemOpen
  \bibfield  {author} {\bibinfo {author} {\bibfnamefont {A.}~\bibnamefont
  {Mook}}, \bibinfo {author} {\bibfnamefont {J.}~\bibnamefont {Henk}}, \ and\
  \bibinfo {author} {\bibfnamefont {I.}~\bibnamefont {Mertig}},\ }\bibfield
  {title} {\enquote {\bibinfo {title} {Edge states in topological magnon
  insulators},}\ }\href {\doibase 10.1103/PhysRevB.90.024412} {\bibfield
  {journal} {\bibinfo  {journal} {Phys. Rev. B}\ }\textbf {\bibinfo {volume}
  {90}},\ \bibinfo {pages} {024412} (\bibinfo {year} {2014})}\BibitemShut
  {NoStop}%
\bibitem [{\citenamefont {Wang}\ \emph {et~al.}(2018)\citenamefont {Wang},
  \citenamefont {Zhang},\ and\ \citenamefont {Wang}}]{Wang2018}%
  \BibitemOpen
  \bibfield  {author} {\bibinfo {author} {\bibfnamefont {X.~S.}\ \bibnamefont
  {Wang}}, \bibinfo {author} {\bibfnamefont {H.~W.}\ \bibnamefont {Zhang}}, \
  and\ \bibinfo {author} {\bibfnamefont {X.~R.}\ \bibnamefont {Wang}},\
  }\bibfield  {title} {\enquote {\bibinfo {title} {Topological magnonics: A
  paradigm for spin-wave manipulation and device design},}\ }\href {\doibase
  10.1103/PhysRevApplied.9.024029} {\bibfield  {journal} {\bibinfo  {journal}
  {Phys. Rev. Applied}\ }\textbf {\bibinfo {volume} {9}},\ \bibinfo {pages}
  {024029} (\bibinfo {year} {2018})}\BibitemShut {NoStop}%
\bibitem [{\citenamefont {D\'{\i}az}\ \emph {et~al.}(2019)\citenamefont
  {D\'{\i}az}, \citenamefont {Klinovaja},\ and\ \citenamefont
  {Loss}}]{Diaz2019}%
  \BibitemOpen
  \bibfield  {author} {\bibinfo {author} {\bibfnamefont {S.~A.}\ \bibnamefont
  {D\'{\i}az}}, \bibinfo {author} {\bibfnamefont {J.}~\bibnamefont
  {Klinovaja}}, \ and\ \bibinfo {author} {\bibfnamefont {D.}~\bibnamefont
  {Loss}},\ }\bibfield  {title} {\enquote {\bibinfo {title} {Topological
  magnons and edge states in antiferromagnetic skyrmion crystals},}\ }\href
  {\doibase 10.1103/PhysRevLett.122.187203} {\bibfield  {journal} {\bibinfo
  {journal} {Phys. Rev. Lett.}\ }\textbf {\bibinfo {volume} {122}},\ \bibinfo
  {pages} {187203} (\bibinfo {year} {2019})}\BibitemShut {NoStop}%
\bibitem [{\citenamefont {Malki}\ and\ \citenamefont
  {Uhrig}(2019)}]{Malki2019}%
  \BibitemOpen
  \bibfield  {author} {\bibinfo {author} {\bibfnamefont {M.}~\bibnamefont
  {Malki}}\ and\ \bibinfo {author} {\bibfnamefont {G.~S.}\ \bibnamefont
  {Uhrig}},\ }\bibfield  {title} {\enquote {\bibinfo {title} {Topological
  magnon bands for magnonics},}\ }\href {\doibase 10.1103/PhysRevB.99.174412}
  {\bibfield  {journal} {\bibinfo  {journal} {Phys. Rev. B}\ }\textbf {\bibinfo
  {volume} {99}},\ \bibinfo {pages} {174412} (\bibinfo {year}
  {2019})}\BibitemShut {NoStop}%
\bibitem [{\citenamefont {M{\"u}hlbauer}\ \emph {et~al.}(2009)\citenamefont
  {M{\"u}hlbauer}, \citenamefont {Binz}, \citenamefont {Jonietz}, \citenamefont
  {Pfleiderer}, \citenamefont {Rosch}, \citenamefont {Neubauer}, \citenamefont
  {Georgii},\ and\ \citenamefont {B{\"o}ni}}]{Muhlbauer2009}%
  \BibitemOpen
  \bibfield  {author} {\bibinfo {author} {\bibfnamefont {S.}~\bibnamefont
  {M{\"u}hlbauer}}, \bibinfo {author} {\bibfnamefont {B.}~\bibnamefont {Binz}},
  \bibinfo {author} {\bibfnamefont {F.}~\bibnamefont {Jonietz}}, \bibinfo
  {author} {\bibfnamefont {C.}~\bibnamefont {Pfleiderer}}, \bibinfo {author}
  {\bibfnamefont {A.}~\bibnamefont {Rosch}}, \bibinfo {author} {\bibfnamefont
  {A.}~\bibnamefont {Neubauer}}, \bibinfo {author} {\bibfnamefont
  {R.}~\bibnamefont {Georgii}}, \ and\ \bibinfo {author} {\bibfnamefont
  {P.}~\bibnamefont {B{\"o}ni}},\ }\bibfield  {title} {\enquote {\bibinfo
  {title} {Skyrmion lattice in a chiral magnet},}\ }\href {\doibase
  10.1126/science.1166767} {\bibfield  {journal} {\bibinfo  {journal}
  {Science}\ }\textbf {\bibinfo {volume} {323}},\ \bibinfo {pages} {915}
  (\bibinfo {year} {2009})}\BibitemShut {NoStop}%
\bibitem [{\citenamefont {Yu}\ \emph {et~al.}(2010)\citenamefont {Yu},
  \citenamefont {Onose}, \citenamefont {Kanazawa}, \citenamefont {Park},
  \citenamefont {Han}, \citenamefont {Matsui}, \citenamefont {Nagaosa},\ and\
  \citenamefont {Tokura}}]{Yu2010}%
  \BibitemOpen
  \bibfield  {author} {\bibinfo {author} {\bibfnamefont {X.~Z.}\ \bibnamefont
  {Yu}}, \bibinfo {author} {\bibfnamefont {Y.}~\bibnamefont {Onose}}, \bibinfo
  {author} {\bibfnamefont {N.}~\bibnamefont {Kanazawa}}, \bibinfo {author}
  {\bibfnamefont {J.~H.}\ \bibnamefont {Park}}, \bibinfo {author}
  {\bibfnamefont {J.~H.}\ \bibnamefont {Han}}, \bibinfo {author} {\bibfnamefont
  {Y.}~\bibnamefont {Matsui}}, \bibinfo {author} {\bibfnamefont
  {N.}~\bibnamefont {Nagaosa}}, \ and\ \bibinfo {author} {\bibfnamefont
  {Y.}~\bibnamefont {Tokura}},\ }\bibfield  {title} {\enquote {\bibinfo {title}
  {Real-space observation of a two-dimensional skyrmion crystal},}\ }\href
  {http://dx.doi.org/10.1038/nature09124} {\bibfield  {journal} {\bibinfo
  {journal} {Nature}\ }\textbf {\bibinfo {volume} {465}},\ \bibinfo {pages}
  {901} (\bibinfo {year} {2010})}\BibitemShut {NoStop}%
\bibitem [{\citenamefont {Zang}\ \emph {et~al.}(2011)\citenamefont {Zang},
  \citenamefont {Mostovoy}, \citenamefont {Han},\ and\ \citenamefont
  {Nagaosa}}]{Zang2011}%
  \BibitemOpen
  \bibfield  {author} {\bibinfo {author} {\bibfnamefont {J.}~\bibnamefont
  {Zang}}, \bibinfo {author} {\bibfnamefont {M.}~\bibnamefont {Mostovoy}},
  \bibinfo {author} {\bibfnamefont {J.~H.}\ \bibnamefont {Han}}, \ and\
  \bibinfo {author} {\bibfnamefont {N.}~\bibnamefont {Nagaosa}},\ }\bibfield
  {title} {\enquote {\bibinfo {title} {Dynamics of skyrmion crystals in
  metallic thin films},}\ }\href {\doibase 10.1103/PhysRevLett.107.136804}
  {\bibfield  {journal} {\bibinfo  {journal} {Phys. Rev. Lett.}\ }\textbf
  {\bibinfo {volume} {107}},\ \bibinfo {pages} {136804} (\bibinfo {year}
  {2011})}\BibitemShut {NoStop}%
\bibitem [{\citenamefont {Petrova}\ and\ \citenamefont
  {Tchernyshyov}(2011)}]{Petrova2011}%
  \BibitemOpen
  \bibfield  {author} {\bibinfo {author} {\bibfnamefont {O.}~\bibnamefont
  {Petrova}}\ and\ \bibinfo {author} {\bibfnamefont {O.}~\bibnamefont
  {Tchernyshyov}},\ }\bibfield  {title} {\enquote {\bibinfo {title} {Spin waves
  in a skyrmion crystal},}\ }\href {\doibase 10.1103/PhysRevB.84.214433}
  {\bibfield  {journal} {\bibinfo  {journal} {Phys. Rev. B}\ }\textbf {\bibinfo
  {volume} {84}},\ \bibinfo {pages} {214433} (\bibinfo {year}
  {2011})}\BibitemShut {NoStop}%
\bibitem [{\citenamefont {Mochizuki}(2012)}]{Mochizuki2012}%
  \BibitemOpen
  \bibfield  {author} {\bibinfo {author} {\bibfnamefont {M.}~\bibnamefont
  {Mochizuki}},\ }\bibfield  {title} {\enquote {\bibinfo {title} {Spin-wave
  modes and their intense excitation effects in skyrmion crystals},}\ }\href
  {\doibase 10.1103/PhysRevLett.108.017601} {\bibfield  {journal} {\bibinfo
  {journal} {Phys. Rev. Lett.}\ }\textbf {\bibinfo {volume} {108}},\ \bibinfo
  {pages} {017601} (\bibinfo {year} {2012})}\BibitemShut {NoStop}%
\bibitem [{\citenamefont {Okamura}\ \emph {et~al.}(2013)\citenamefont
  {Okamura}, \citenamefont {Kagawa}, \citenamefont {Mochizuki}, \citenamefont
  {Kubota}, \citenamefont {Seki}, \citenamefont {Ishiwata}, \citenamefont
  {Kawasaki}, \citenamefont {Onose},\ and\ \citenamefont
  {Tokura}}]{Okamura2013}%
  \BibitemOpen
  \bibfield  {author} {\bibinfo {author} {\bibfnamefont {Y.}~\bibnamefont
  {Okamura}}, \bibinfo {author} {\bibfnamefont {F.}~\bibnamefont {Kagawa}},
  \bibinfo {author} {\bibfnamefont {M.}~\bibnamefont {Mochizuki}}, \bibinfo
  {author} {\bibfnamefont {M.}~\bibnamefont {Kubota}}, \bibinfo {author}
  {\bibfnamefont {S.}~\bibnamefont {Seki}}, \bibinfo {author} {\bibfnamefont
  {S.}~\bibnamefont {Ishiwata}}, \bibinfo {author} {\bibfnamefont
  {M.}~\bibnamefont {Kawasaki}}, \bibinfo {author} {\bibfnamefont
  {Y.}~\bibnamefont {Onose}}, \ and\ \bibinfo {author} {\bibfnamefont
  {Y.}~\bibnamefont {Tokura}},\ }\bibfield  {title} {\enquote {\bibinfo {title}
  {Microwave magnetoelectric effect via skyrmion resonance modes in a
  helimagnetic multiferroic},}\ }\href {\doibase 10.1038/ncomms3391} {\bibfield
   {journal} {\bibinfo  {journal} {Nature Communications}\ }\textbf {\bibinfo
  {volume} {4}},\ \bibinfo {pages} {2391} (\bibinfo {year} {2013})}\BibitemShut
  {NoStop}%
\bibitem [{\citenamefont {Rold{\'a}n-Molina}\ \emph {et~al.}(2016)\citenamefont
  {Rold{\'a}n-Molina}, \citenamefont {N\'u\~nez},\ and\ \citenamefont
  {Fern{\'a}ndez-Rossier}}]{Roldan-Molina2016}%
  \BibitemOpen
  \bibfield  {author} {\bibinfo {author} {\bibfnamefont {A.}~\bibnamefont
  {Rold{\'a}n-Molina}}, \bibinfo {author} {\bibfnamefont {A.~S.}\ \bibnamefont
  {N\'u\~nez}}, \ and\ \bibinfo {author} {\bibfnamefont {J.}~\bibnamefont
  {Fern{\'a}ndez-Rossier}},\ }\bibfield  {title} {\enquote {\bibinfo {title}
  {Topological spin waves in the atomic-scale magnetic skyrmion crystal},}\
  }\href {http://stacks.iop.org/1367-2630/18/i=4/a=045015} {\bibfield
  {journal} {\bibinfo  {journal} {New J. Phys.}\ }\textbf {\bibinfo {volume}
  {18}},\ \bibinfo {pages} {045015} (\bibinfo {year} {2016})}\BibitemShut
  {NoStop}%
\bibitem [{\citenamefont {Evans}\ \emph {et~al.}(2014)\citenamefont {Evans},
  \citenamefont {Fan}, \citenamefont {Chureemart}, \citenamefont {Ostler},
  \citenamefont {Ellis},\ and\ \citenamefont {Chantrell}}]{Evans2014}%
  \BibitemOpen
  \bibfield  {author} {\bibinfo {author} {\bibfnamefont {R.~F.~L.}\
  \bibnamefont {Evans}}, \bibinfo {author} {\bibfnamefont {W.~J.}\ \bibnamefont
  {Fan}}, \bibinfo {author} {\bibfnamefont {P.}~\bibnamefont {Chureemart}},
  \bibinfo {author} {\bibfnamefont {T.~A.}\ \bibnamefont {Ostler}}, \bibinfo
  {author} {\bibfnamefont {M.~O.~A.}\ \bibnamefont {Ellis}}, \ and\ \bibinfo
  {author} {\bibfnamefont {R.~W.}\ \bibnamefont {Chantrell}},\ }\bibfield
  {title} {\enquote {\bibinfo {title} {Atomistic spin model simulations of
  magnetic nanomaterials},}\ }\href
  {http://stacks.iop.org/0953-8984/26/i=10/a=103202} {\bibfield  {journal}
  {\bibinfo  {journal} {J. Phys.: Condens. Matter}\ }\textbf {\bibinfo {volume}
  {26}},\ \bibinfo {pages} {103202} (\bibinfo {year} {2014})}\BibitemShut
  {NoStop}%
\bibitem [{\citenamefont {Berkov}\ and\ \citenamefont
  {Gorn}(2006)}]{Berkov2005}%
  \BibitemOpen
  \bibfield  {author} {\bibinfo {author} {\bibfnamefont {D.~V.}\ \bibnamefont
  {Berkov}}\ and\ \bibinfo {author} {\bibfnamefont {N.~L.}\ \bibnamefont
  {Gorn}},\ }\enquote {\bibinfo {title} {Handbook of advanced magnetic
  materials: Vol 2. characterization and simulation},}\ \ (\bibinfo
  {publisher} {Springer, Boston},\ \bibinfo {year} {2006})\ p.\ \bibinfo
  {pages} {447}\BibitemShut {NoStop}%
\bibitem [{\citenamefont {Holstein}\ and\ \citenamefont
  {Primakoff}(1940)}]{Holstein1940}%
  \BibitemOpen
  \bibfield  {author} {\bibinfo {author} {\bibfnamefont {T.}~\bibnamefont
  {Holstein}}\ and\ \bibinfo {author} {\bibfnamefont {H.}~\bibnamefont
  {Primakoff}},\ }\bibfield  {title} {\enquote {\bibinfo {title} {Field
  dependence of the intrinsic domain magnetization of a ferromagnet},}\ }\href
  {\doibase 10.1103/PhysRev.58.1098} {\bibfield  {journal} {\bibinfo  {journal}
  {Phys. Rev.}\ }\textbf {\bibinfo {volume} {58}},\ \bibinfo {pages} {1098}
  (\bibinfo {year} {1940})}\BibitemShut {NoStop}%
\bibitem [{\citenamefont {Kittel}(1963)}]{Kittel1963}%
  \BibitemOpen
  \bibfield  {author} {\bibinfo {author} {\bibfnamefont {C.}~\bibnamefont
  {Kittel}},\ }\href@noop {} {\emph {\bibinfo {title} {Quantum Theory of
  Solids}}}\ (\bibinfo  {publisher} {Wiley, New York},\ \bibinfo {year}
  {1963})\BibitemShut {NoStop}%
\bibitem [{\citenamefont {Lin}\ \emph {et~al.}(2014)\citenamefont {Lin},
  \citenamefont {Batista},\ and\ \citenamefont {Saxena}}]{Lin2014}%
  \BibitemOpen
  \bibfield  {author} {\bibinfo {author} {\bibfnamefont {S.-Z.}\ \bibnamefont
  {Lin}}, \bibinfo {author} {\bibfnamefont {C.~D.}\ \bibnamefont {Batista}}, \
  and\ \bibinfo {author} {\bibfnamefont {A.}~\bibnamefont {Saxena}},\
  }\bibfield  {title} {\enquote {\bibinfo {title} {Internal modes of a skyrmion
  in the ferromagnetic state of chiral magnets},}\ }\href {\doibase
  10.1103/PhysRevB.89.024415} {\bibfield  {journal} {\bibinfo  {journal} {Phys.
  Rev. B}\ }\textbf {\bibinfo {volume} {89}},\ \bibinfo {pages} {024415}
  (\bibinfo {year} {2014})}\BibitemShut {NoStop}%
\bibitem [{\citenamefont {Psaroudaki}\ and\ \citenamefont
  {Loss}(2018)}]{Psaroudaki2018}%
  \BibitemOpen
  \bibfield  {author} {\bibinfo {author} {\bibfnamefont {C.}~\bibnamefont
  {Psaroudaki}}\ and\ \bibinfo {author} {\bibfnamefont {D.}~\bibnamefont
  {Loss}},\ }\bibfield  {title} {\enquote {\bibinfo {title} {Skyrmions driven
  by intrinsic magnons},}\ }\href {\doibase 10.1103/PhysRevLett.120.237203}
  {\bibfield  {journal} {\bibinfo  {journal} {Phys. Rev. Lett.}\ }\textbf
  {\bibinfo {volume} {120}},\ \bibinfo {pages} {237203} (\bibinfo {year}
  {2018})}\BibitemShut {NoStop}%
\bibitem [{\citenamefont {Onose}\ \emph {et~al.}(2012)\citenamefont {Onose},
  \citenamefont {Okamura}, \citenamefont {Seki}, \citenamefont {Ishiwata},\
  and\ \citenamefont {Tokura}}]{Onose2012}%
  \BibitemOpen
  \bibfield  {author} {\bibinfo {author} {\bibfnamefont {Y.}~\bibnamefont
  {Onose}}, \bibinfo {author} {\bibfnamefont {Y.}~\bibnamefont {Okamura}},
  \bibinfo {author} {\bibfnamefont {S.}~\bibnamefont {Seki}}, \bibinfo {author}
  {\bibfnamefont {S.}~\bibnamefont {Ishiwata}}, \ and\ \bibinfo {author}
  {\bibfnamefont {Y.}~\bibnamefont {Tokura}},\ }\bibfield  {title} {\enquote
  {\bibinfo {title} {Observation of magnetic excitations of skyrmion crystal in
  a helimagnetic insulator {Cu$_{2}$OSeO$_{3}$}},}\ }\href {\doibase
  10.1103/PhysRevLett.109.037603} {\bibfield  {journal} {\bibinfo  {journal}
  {Phys. Rev. Lett.}\ }\textbf {\bibinfo {volume} {109}},\ \bibinfo {pages}
  {037603} (\bibinfo {year} {2012})}\BibitemShut {NoStop}%
\bibitem [{\citenamefont {Schwarze}\ \emph {et~al.}(2015)\citenamefont
  {Schwarze}, \citenamefont {Waizner}, \citenamefont {Garst}, \citenamefont
  {Bauer}, \citenamefont {Stasinopoulos}, \citenamefont {Berger}, \citenamefont
  {Pfleiderer},\ and\ \citenamefont {Grundler}}]{Schwarze2015}%
  \BibitemOpen
  \bibfield  {author} {\bibinfo {author} {\bibfnamefont {T.}~\bibnamefont
  {Schwarze}}, \bibinfo {author} {\bibfnamefont {J.}~\bibnamefont {Waizner}},
  \bibinfo {author} {\bibfnamefont {M.}~\bibnamefont {Garst}}, \bibinfo
  {author} {\bibfnamefont {A.}~\bibnamefont {Bauer}}, \bibinfo {author}
  {\bibfnamefont {I.}~\bibnamefont {Stasinopoulos}}, \bibinfo {author}
  {\bibfnamefont {H.}~\bibnamefont {Berger}}, \bibinfo {author} {\bibfnamefont
  {C.}~\bibnamefont {Pfleiderer}}, \ and\ \bibinfo {author} {\bibfnamefont
  {D.}~\bibnamefont {Grundler}},\ }\bibfield  {title} {\enquote {\bibinfo
  {title} {Universal helimagnon and skyrmion excitations in metallic,
  semiconducting and insulating chiral magnets},}\ }\href {\doibase
  10.1038/nmat4223} {\bibfield  {journal} {\bibinfo  {journal} {Nature
  Materials}\ }\textbf {\bibinfo {volume} {14}},\ \bibinfo {pages} {478}
  (\bibinfo {year} {2015})}\BibitemShut {NoStop}%
\bibitem [{\citenamefont {Ehlers}\ \emph {et~al.}(2016)\citenamefont {Ehlers},
  \citenamefont {Stasinopoulos}, \citenamefont {Tsurkan}, \citenamefont
  {Krug~von Nidda}, \citenamefont {Feh\'er}, \citenamefont {Leonov},
  \citenamefont {K\'ezsm\'arki}, \citenamefont {Grundler},\ and\ \citenamefont
  {Loidl}}]{Ehlers2016}%
  \BibitemOpen
  \bibfield  {author} {\bibinfo {author} {\bibfnamefont {D.}~\bibnamefont
  {Ehlers}}, \bibinfo {author} {\bibfnamefont {I.}~\bibnamefont
  {Stasinopoulos}}, \bibinfo {author} {\bibfnamefont {V.}~\bibnamefont
  {Tsurkan}}, \bibinfo {author} {\bibfnamefont {H.-A.}\ \bibnamefont {Krug~von
  Nidda}}, \bibinfo {author} {\bibfnamefont {T.}~\bibnamefont {Feh\'er}},
  \bibinfo {author} {\bibfnamefont {A.}~\bibnamefont {Leonov}}, \bibinfo
  {author} {\bibfnamefont {I.}~\bibnamefont {K\'ezsm\'arki}}, \bibinfo {author}
  {\bibfnamefont {D.}~\bibnamefont {Grundler}}, \ and\ \bibinfo {author}
  {\bibfnamefont {A.}~\bibnamefont {Loidl}},\ }\bibfield  {title} {\enquote
  {\bibinfo {title} {Skyrmion dynamics under uniaxial anisotropy},}\ }\href
  {\doibase 10.1103/PhysRevB.94.014406} {\bibfield  {journal} {\bibinfo
  {journal} {Phys. Rev. B}\ }\textbf {\bibinfo {volume} {94}},\ \bibinfo
  {pages} {014406} (\bibinfo {year} {2016})}\BibitemShut {NoStop}%
\bibitem [{\citenamefont {Bansil}\ \emph {et~al.}(2016)\citenamefont {Bansil},
  \citenamefont {Lin},\ and\ \citenamefont {Das}}]{Bansil2016}%
  \BibitemOpen
  \bibfield  {author} {\bibinfo {author} {\bibfnamefont {A.}~\bibnamefont
  {Bansil}}, \bibinfo {author} {\bibfnamefont {H.}~\bibnamefont {Lin}}, \ and\
  \bibinfo {author} {\bibfnamefont {T.}~\bibnamefont {Das}},\ }\bibfield
  {title} {\enquote {\bibinfo {title} {Colloquium: Topological band theory},}\
  }\href {\doibase 10.1103/RevModPhys.88.021004} {\bibfield  {journal}
  {\bibinfo  {journal} {Rev. Mod. Phys.}\ }\textbf {\bibinfo {volume} {88}},\
  \bibinfo {pages} {021004} (\bibinfo {year} {2016})}\BibitemShut {NoStop}%
\bibitem [{\citenamefont {Hatsugai}(1993{\natexlab{a}})}]{Hatsugai1993a}%
  \BibitemOpen
  \bibfield  {author} {\bibinfo {author} {\bibfnamefont {Y.}~\bibnamefont
  {Hatsugai}},\ }\bibfield  {title} {\enquote {\bibinfo {title} {Edge states in
  the integer quantum hall effect and the riemann surface of the bloch
  function},}\ }\href {\doibase 10.1103/PhysRevB.48.11851} {\bibfield
  {journal} {\bibinfo  {journal} {Phys. Rev. B}\ }\textbf {\bibinfo {volume}
  {48}},\ \bibinfo {pages} {11851} (\bibinfo {year}
  {1993}{\natexlab{a}})}\BibitemShut {NoStop}%
\bibitem [{\citenamefont {Hatsugai}(1993{\natexlab{b}})}]{Hatsugai1993b}%
  \BibitemOpen
  \bibfield  {author} {\bibinfo {author} {\bibfnamefont {Y.}~\bibnamefont
  {Hatsugai}},\ }\bibfield  {title} {\enquote {\bibinfo {title} {Chern number
  and edge states in the integer quantum hall effect},}\ }\href {\doibase
  10.1103/PhysRevLett.71.3697} {\bibfield  {journal} {\bibinfo  {journal}
  {Phys. Rev. Lett.}\ }\textbf {\bibinfo {volume} {71}},\ \bibinfo {pages}
  {3697} (\bibinfo {year} {1993}{\natexlab{b}})}\BibitemShut {NoStop}%
\bibitem [{\citenamefont {Rhim}\ \emph {et~al.}(2018)\citenamefont {Rhim},
  \citenamefont {Bardarson},\ and\ \citenamefont {Slager}}]{Rhim2018}%
  \BibitemOpen
  \bibfield  {author} {\bibinfo {author} {\bibfnamefont {J.-W.}\ \bibnamefont
  {Rhim}}, \bibinfo {author} {\bibfnamefont {J.~H.}\ \bibnamefont {Bardarson}},
  \ and\ \bibinfo {author} {\bibfnamefont {R.-J.}\ \bibnamefont {Slager}},\
  }\bibfield  {title} {\enquote {\bibinfo {title} {Unified bulk-boundary
  correspondence for band insulators},}\ }\href {\doibase
  10.1103/PhysRevB.97.115143} {\bibfield  {journal} {\bibinfo  {journal} {Phys.
  Rev. B}\ }\textbf {\bibinfo {volume} {97}},\ \bibinfo {pages} {115143}
  (\bibinfo {year} {2018})}\BibitemShut {NoStop}%
\bibitem [{\citenamefont {R{\"o}{\ss}ler}\ \emph {et~al.}(2011)\citenamefont
  {R{\"o}{\ss}ler}, \citenamefont {Leonov},\ and\ \citenamefont
  {Bogdanov}}]{Roessler2011}%
  \BibitemOpen
  \bibfield  {author} {\bibinfo {author} {\bibfnamefont {U.~K.}\ \bibnamefont
  {R{\"o}{\ss}ler}}, \bibinfo {author} {\bibfnamefont {A.~A.}\ \bibnamefont
  {Leonov}}, \ and\ \bibinfo {author} {\bibfnamefont {A.~N.}\ \bibnamefont
  {Bogdanov}},\ }\bibfield  {title} {\enquote {\bibinfo {title} {Chiral
  skyrmionic matter in non-centrosymmetric magnets},}\ }\href {\doibase
  10.1088/1742-6596/303/1/012105} {\bibfield  {journal} {\bibinfo  {journal}
  {Journal of Physics: Conference Series}\ }\textbf {\bibinfo {volume} {303}},\
  \bibinfo {pages} {012105} (\bibinfo {year} {2011})}\BibitemShut {NoStop}%
\bibitem [{\citenamefont {Brataas}\ \emph {et~al.}(2002)\citenamefont
  {Brataas}, \citenamefont {Tserkovnyak}, \citenamefont {Bauer},\ and\
  \citenamefont {Halperin}}]{Brataas2002}%
  \BibitemOpen
  \bibfield  {author} {\bibinfo {author} {\bibfnamefont {A.}~\bibnamefont
  {Brataas}}, \bibinfo {author} {\bibfnamefont {Y.}~\bibnamefont
  {Tserkovnyak}}, \bibinfo {author} {\bibfnamefont {G.~E.~W.}\ \bibnamefont
  {Bauer}}, \ and\ \bibinfo {author} {\bibfnamefont {B.~I.}\ \bibnamefont
  {Halperin}},\ }\bibfield  {title} {\enquote {\bibinfo {title} {Spin battery
  operated by ferromagnetic resonance},}\ }\href {\doibase
  10.1103/PhysRevB.66.060404} {\bibfield  {journal} {\bibinfo  {journal} {Phys.
  Rev. B}\ }\textbf {\bibinfo {volume} {66}},\ \bibinfo {pages} {060404}
  (\bibinfo {year} {2002})}\BibitemShut {NoStop}%
\bibitem [{\citenamefont {Du}\ \emph {et~al.}(2017)\citenamefont {Du},
  \citenamefont {van~der Sar}, \citenamefont {Zhou}, \citenamefont {Upadhyaya},
  \citenamefont {Casola}, \citenamefont {Zhang}, \citenamefont {Onbasli},
  \citenamefont {Ross}, \citenamefont {Walsworth}, \citenamefont
  {Tserkovnyak},\ and\ \citenamefont {Yacoby}}]{Du2017}%
  \BibitemOpen
  \bibfield  {author} {\bibinfo {author} {\bibfnamefont {C.}~\bibnamefont
  {Du}}, \bibinfo {author} {\bibfnamefont {T.}~\bibnamefont {van~der Sar}},
  \bibinfo {author} {\bibfnamefont {T.~X.}\ \bibnamefont {Zhou}}, \bibinfo
  {author} {\bibfnamefont {P.}~\bibnamefont {Upadhyaya}}, \bibinfo {author}
  {\bibfnamefont {F.}~\bibnamefont {Casola}}, \bibinfo {author} {\bibfnamefont
  {H.}~\bibnamefont {Zhang}}, \bibinfo {author} {\bibfnamefont {M.~C.}\
  \bibnamefont {Onbasli}}, \bibinfo {author} {\bibfnamefont {C.~A.}\
  \bibnamefont {Ross}}, \bibinfo {author} {\bibfnamefont {R.~L.}\ \bibnamefont
  {Walsworth}}, \bibinfo {author} {\bibfnamefont {Y.}~\bibnamefont
  {Tserkovnyak}}, \ and\ \bibinfo {author} {\bibfnamefont {A.}~\bibnamefont
  {Yacoby}},\ }\bibfield  {title} {\enquote {\bibinfo {title} {Control and
  local measurement of the spin chemical potential in a magnetic insulator},}\
  }\href {\doibase 10.1126/science.aak9611} {\bibfield  {journal} {\bibinfo
  {journal} {Science}\ }\textbf {\bibinfo {volume} {357}},\ \bibinfo {pages}
  {195} (\bibinfo {year} {2017})}\BibitemShut {NoStop}%
\bibitem [{\citenamefont {Demokritov}\ \emph {et~al.}(2001)\citenamefont
  {Demokritov}, \citenamefont {Hillebrands},\ and\ \citenamefont
  {Slavin}}]{Demokritov2001}%
  \BibitemOpen
  \bibfield  {author} {\bibinfo {author} {\bibfnamefont {S.}~\bibnamefont
  {Demokritov}}, \bibinfo {author} {\bibfnamefont {B.}~\bibnamefont
  {Hillebrands}}, \ and\ \bibinfo {author} {\bibfnamefont {A.}~\bibnamefont
  {Slavin}},\ }\bibfield  {title} {\enquote {\bibinfo {title} {Brillouin light
  scattering studies of confined spin waves: linear and nonlinear
  confinement},}\ }\href {\doibase 10.1016/S0370-1573(00)00116-2} {\bibfield
  {journal} {\bibinfo  {journal} {Phys. Rep.}\ }\textbf {\bibinfo {volume}
  {348}},\ \bibinfo {pages} {441 } (\bibinfo {year} {2001})}\BibitemShut
  {NoStop}%
\bibitem [{\citenamefont {Zagury}\ and\ \citenamefont
  {Rezende}(1971)}]{Zagury1971}%
  \BibitemOpen
  \bibfield  {author} {\bibinfo {author} {\bibfnamefont {N.}~\bibnamefont
  {Zagury}}\ and\ \bibinfo {author} {\bibfnamefont {S.~M.}\ \bibnamefont
  {Rezende}},\ }\bibfield  {title} {\enquote {\bibinfo {title} {Theory of
  macroscopic excitations of magnons},}\ }\href {\doibase
  10.1103/PhysRevB.4.201} {\bibfield  {journal} {\bibinfo  {journal} {Phys.
  Rev. B}\ }\textbf {\bibinfo {volume} {4}},\ \bibinfo {pages} {201} (\bibinfo
  {year} {1971})}\BibitemShut {NoStop}%
\bibitem [{\citenamefont {Majlis}(2007)}]{Majlis2007}%
  \BibitemOpen
  \bibfield  {author} {\bibinfo {author} {\bibfnamefont {N.}~\bibnamefont
  {Majlis}},\ }\href {\doibase 10.1142/6094} {\emph {\bibinfo {title} {The
  Quantum Theory of Magnetism}}},\ \bibinfo {edition} {2nd}\ ed.\ (\bibinfo
  {publisher} {WORLD SCIENTIFIC},\ \bibinfo {year} {2007})\BibitemShut
  {NoStop}%
\end{thebibliography}
\end{document}